\documentclass[12pt]{iopart}
\usepackage[pdftex]{graphicx}

\usepackage{iopams}  
\usepackage{amsthm}
 \newtheorem{theo}{Theorem}

\begin{document}

\title[]{Two extensions of exact non-equilibrium steady states of a boundary driven cellular automaton}

\author{Atsuo Inoue and Shinji Takesue}

\address{Department of Physics, Graduate School of Science, Kyoto University,
Kyoto 606-8502, Japan}
\eads{\mailto{inoue.atsuo.26u@st.kyoto-u.ac.jp},\mailto{takesue@scphys.kyoto-u.ac.jp}}
\vspace{10pt}
\begin{indented}
\item[]
\end{indented}

\begin{abstract}
Recently Prosen and Mej\'{i}a-Monasterio 
(\jpa \textbf{49} (2016) 185003)
obtained exact nonequilibrium steady states
of an integrable and reversible cellular automaton driven by some stochastic 
boundary conditions.
In this paper,  we explore the possible extensions of their method by generalizing the 
boundary conditions.  As the result, we find 
two cases where such an extension is possible. 
One is the case where a special condition is satisfied in a 
generalized boundary condition.
The other is obtained by considering a conserved quantity as energy and 
boundaries as heat reservoirs.  
The latter includes the original solution as the special case.
Properties of the both solutions are discussed.
\end{abstract}
\maketitle

\section{Introduction}
A cellular automaton (CA) is a discrete dynamical system composed of regularly
ordered cells.  The state of each cell takes values on a finite set.  
The cells simultaneously 
update their states in discrete time according to a deterministic local rule.  
Since Wolfram's work \cite{Wolfram83}, 
CAs are not only studied in traditional computation theory and mathematics
but also widely applied in various fields of science including
fluid mechanics \cite{Rothman97}, 
reaction-diffusion systems \cite{Hayase98},
and integrable dynamical systems \cite{Tokihiro96}.

One of the authors studied one-dimensional reversible CA of the form
\cite{Takesue87,Takesue89, Takesue90}
\begin{equation}
 x_i^{t+1}=f(x_{i-1}^t,x_i^t,x_{i+1}^t)\oplus x_i^{t-1},
\label{ERCA}
\end{equation} 
where $i$ and $t$ denote integers representing cell and time, respectively,
$x_i^t\in\{0,1\}$ means the state of cell $i$ at time $t$, and $\oplus$ the 
\textit{exclusive OR} operation; $0\oplus 0=1\oplus 1=0$ and $0\oplus 1=1\oplus 0=1$. 
This CA is time-reversal invariant because 
the time-reversed evolution follows the same rule as
\begin{equation}
  x_i^{t-1}=f(x_{i-1}^t,x_i^t,x_{i+1}^t)\oplus x_i^{t+1}.
\label{ERCA-reversed}
\end{equation}
This type of CA is the second-order variant of Wolfram's elementary CA, 
so is called elementary reversible CA (ERCA).  Each rule is referred to as
Wolfram's code $\sum_{x,y,z=0,1}f(x,y,z)2^{4x+2y+z}$ appended by an `R'.
For example, if $f(000)=f(010)=0$ and $f(x,y,z)=1$ for othe configurations, 
the rule is called 250R. 
Due to the discrete nature of CA, the time-reversal invariance readily means the
preservation of phase volume like Liouville's theorem in statistical mechanics.
Thus, if an ERCA has an additive conserved quantity, we can define time-invariant
Gibbs measure by considering the conserved quantity as energy.   
A necessary and sufficient condition for a CA to have an additive conserved quantity
was derived and it turns out that some rules in ERCA certainly have such 
conserved quantities \cite{Hattori90}.  
Thus ergodic properties and phase space structures
are examined for some rules \cite{Takesue87,Takesue89}. 
Moreover, we can attach a heat reservoir to either end 
of the system by devising some stochastic update rule for the cell at the end.
When the reservoirs at the left and right ends have different temperatures, 
there occurs transport of \textit{energy}.  
It was numerically revealed that rule 90R shows ballistic transport, 
while rule 26R shows diffusive motion of energy, 
which leads to the Green-Kubo formula for thermal conductivity \cite{Takesue90}.

Recently, Prosen and  Mej\'{i}a-Monasterio \cite{Prosen16} proposed a similar
reversible cellular-automaton model with stochastic boundary conditions which means
chemical baths for absorbing and emitting particles.  
Their model is based on rule 54 (RCA54) presented by 
Bobenko \etal \cite{Bobenko93}
which is defined on a one-dimensional zigzag chain.  At first glance, ECA54 is different
from ERCA, but it is related to rule 250R in ERCA as we will see afterwards.
Prosen and  Mej\'{i}a-Monasterio not only proved the existence 
and uniqueness of a nonequilibrium steady state (NESS) for RCA54, 
but also explicitly obtained an 
exact solution using a form of matrix product ansatz.  This is the first time that
exact NESS is obtained for nontrivial boundary driven CA models.
It is worth looking for new such solutions for other cases.

In this paper, we explorer the possibility of extending their method by extending the
boundary conditions for the same model as in \cite{Prosen16}.  
The boundary condition employed by Prosen and 
Mej\'{i}a-Monasterio is rather special.  We find two generalizatons where such
extension is possible.  One is the case where parameters satisfy a special relation
in generalized boundary conditions and the other is obtained by employing 
the boundary conditions similar to that for ERCA. 
We explain the former in the following section and the latter in 
Section 3.  Section 4 is devoted to summary and discussion.

\section{Boundary driven cellular automaton model}
\label{chapter-RCA54}

\subsection{Definitions}

Though our model is the same as that in \cite{Prosen16} 
except the boundary conditions,
we give a detailed account on the model to make the manuscript self-contained.

\begin{figure}[ht]
\begin{center}
\includegraphics[scale=0.25]{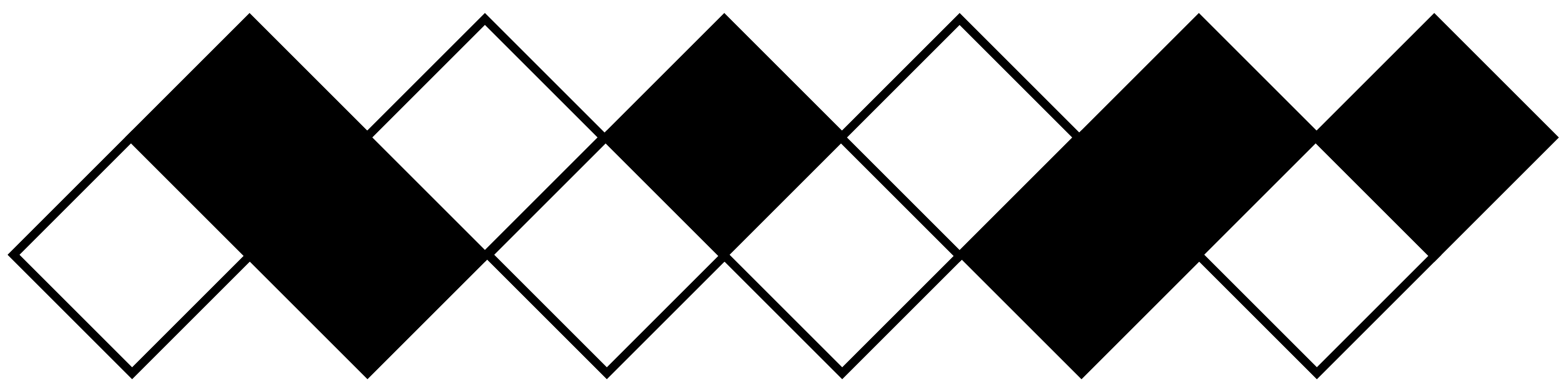}
\end{center}
    \caption{Illustration of a configuration of our system composed of 12 diamond-shaped
cells. Each cell takes value 0 (white) or 1 (black). This figure represents the configuration
 $011001001101$.}
    \label{figure01}
\end{figure}

\begin{figure}[ht]
\begin{center}
\includegraphics[scale=0.25]{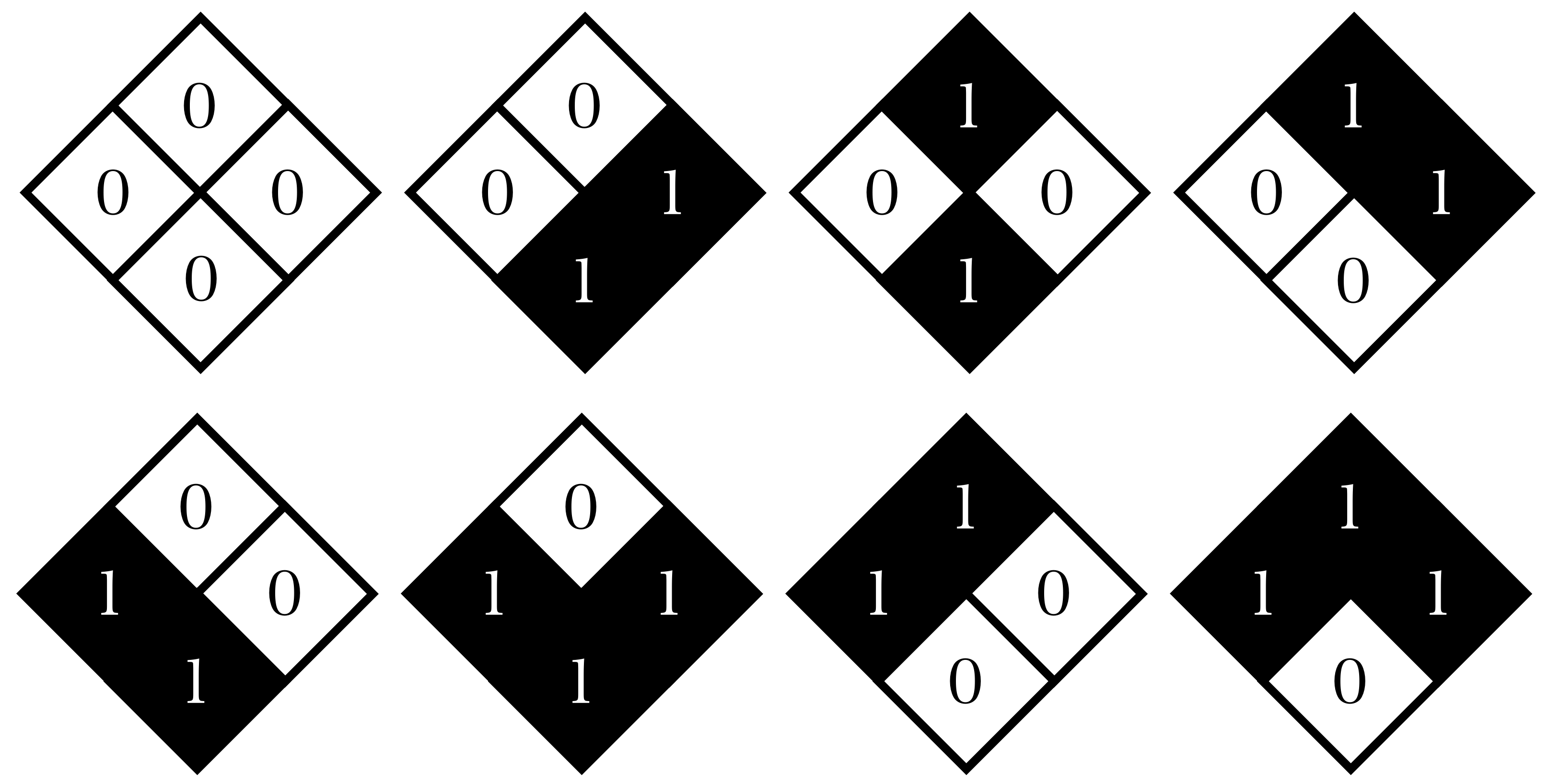}
\end{center}
\caption{All patterns of (\ref{RCA54-def}). 
The value of the bottom cell is determined from the values of the top, left, and right cells.}
    \label{figure03}
\end{figure}

RCA54 is a one-dimensional discrete system which consists of cells connected zigzag 
as in Figure \ref{figure01}. Each cell takes value 0 or 1. 
For simplicity, we assume that the number of cells $n$ is even.  
The value of each cell at the next time step is determined by the following rules:
\begin{eqnarray}
 s^{t+1}_{2k}&=&\chi(s_{2k-1}^t,s_{2k}^t,s_{2k+1}^t) 
\label{CA-rule-even}   \\
  s^{t+1}_{2k+1}&=&\chi(s_{2k}^{t+1},s_{2k+1}^t,s_{2k+2}^{t+1}) , 
\label{CA-rule-odd}
\end{eqnarray}
where $s^t_i\in \mathbb{Z}_2=\{0,1\}$ is a value of the $i$-th cell at time $t$ 
and $\chi:\mathbb{Z}_2\times\mathbb{Z}_2\times\mathbb{Z}_2
\to\mathbb{Z}_2$ 
is defined as
\begin{equation}
 \chi(s,s',s'')=s\oplus s'\oplus s''\oplus ss''.
 \label{RCA54-def}
\end{equation}
This rule is illustrated in  Figure \ref{figure03}.
Equation (\ref{RCA54-def}) has the following property
\begin{equation}
  \chi(s,s',s'')=t\qquad\Leftrightarrow \qquad   \chi(s,t,s'')=s',
  \label{time-reversal-RCA54}
\end{equation}
which means that the CA is time-reversal invariant. 
Wolfram code for this $\chi$ is 54, so this CA is called RCA54.
It should be noticed that there is no relation to Wolfram's elementary CA or
ERCA with the same code number.

\begin{figure}[ht]
\begin{center}
\includegraphics[scale=0.3]{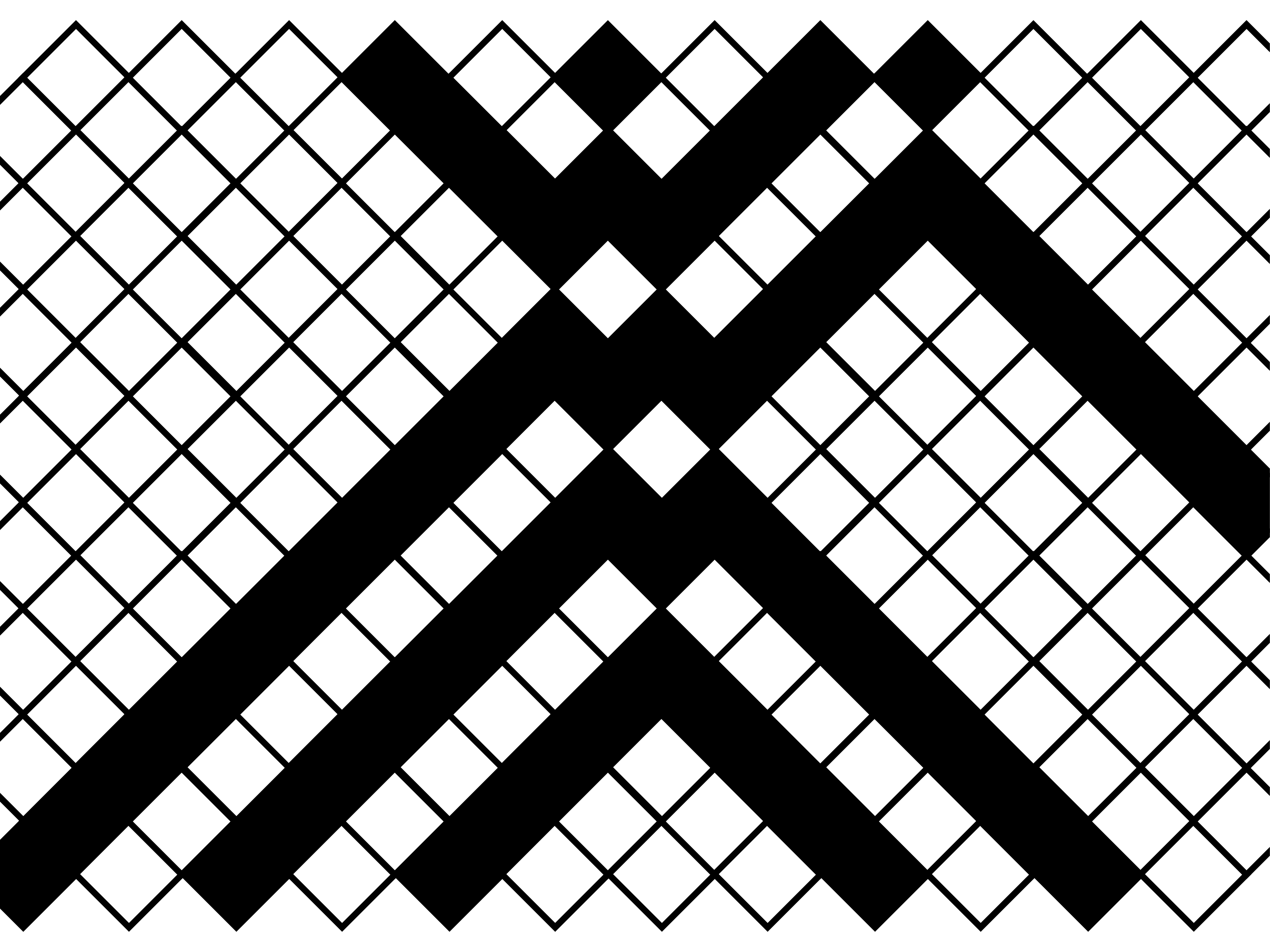}
\end{center}
    \caption{An example of trajectory in RCA54, where the initial configuration is
given by Figure \ref{figure01}. All cells outside the figure are assumed to be 0. 
Value 1 colored black moves to left or right with velocity $\pm 1$ until collision occurs. 
When two particles collide, they shifts to the next lower cell and then move away separately.}
    \label{figure04}
\end{figure}

In RCA54, cells with value 1 look like trajectories of particles moving from side to side 
with velocity $\pm1$ (see Figure \ref{figure04}).  
The time-reversal symmtery (\ref{time-reversal-RCA54}) 
is similar to that of motion in Newtonian mechanics.  

\begin{figure}[ht]
\begin{center}
\includegraphics[scale=0.3]{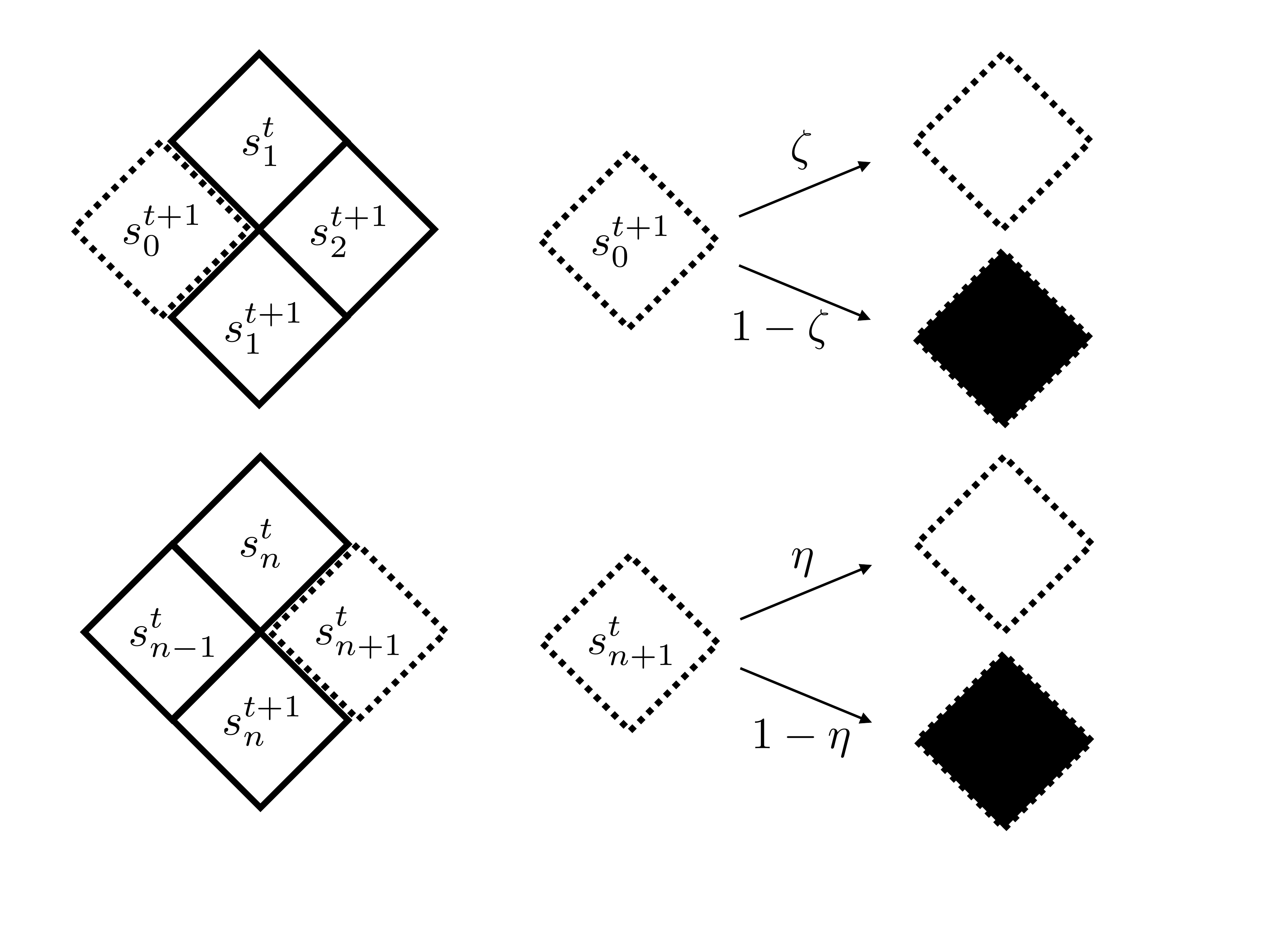}
    \caption{Boundary conditions to give $s^{t+1}_1$ and $s^{t+1}_n$. Make the virtual cell on the outside so that the rule (\ref{RCA54-def}) can be applied. The two parameters are in the range $0 <\zeta, \,\eta <1$.}
    \label{figure05}
    \end{center}
\end{figure}

In finite systems of RCA54, some boundary condition is necessary to determine 
values of the boundaries $i=1$ and $n$, 
which cannot be determined by  (\ref{CA-rule-even}) or (\ref{CA-rule-odd}) 
only. 
For that purpose , we consider virtual cells $i=0$ and $n+1$ outside of 
the both ends of the system. 
Value of the 0th cell is given as $s_0^{t+1}=0$ with probability $\zeta$, 
and $s_0^{t+1}=1$ with $1-\zeta$.  
On the other hand, value of the $(n+1)$-st cell is given as $s_{n+1}^{t+1}=0$ 
with probability $\eta$, and $s_{n+1}^{t+1}=1$ with $1-\eta$. 
Then, we can apply the rule and determine the next time values of both end cells 
(Figure \ref{figure05}).  
This is a generalization from \cite{Prosen16}, where only the case  
$\zeta=\eta=1/2$ is considered.

\begin{figure}[ht]
\begin{center}
\includegraphics[scale=0.3]{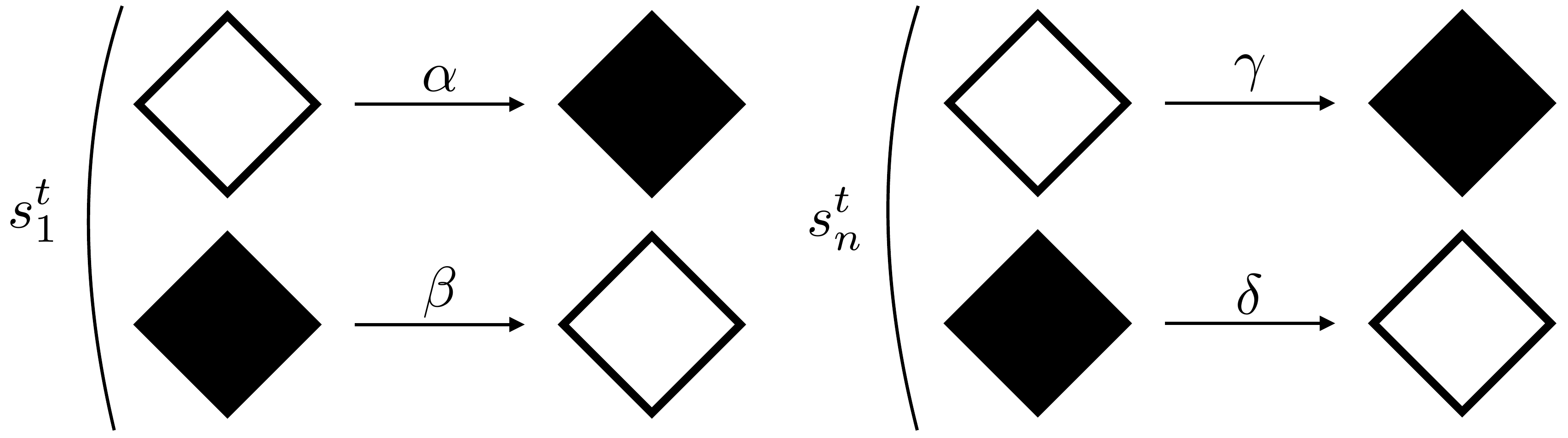}
    \caption{Boundary conditions corresponding to particle absorption/emission, stochastically generate or annihilate the value $1$ at the end cells $ 1, n $. The four parameters are in the range $ 0 <\alpha,\, \beta, \,\gamma,\, \delta <1 $.}
    \label{figure06}
    \end{center}
\end{figure}

Moreover,  we take into account emission and absorption of particles 
at the boundaries as follows. 
If $s_1^t =0$ at time $t$, change it to $ s_1^t = 1$ with probability $\alpha$, 
which corresponds to emission. 
On the other hand, if $s_1^t =1$ at time $t$, change it to $s_1^t=0$ with 
probability $\beta$, 
which corresponds to absorption (Figure \ref{figure06}).

Thus, in RCA54, the bulk is deterministically and the boundary is stochastically 
developed temporally. 
A configuration at time $ t $ is represented by a binary sequence
$\bi{s}^t=s_1^t s_2^t \cdots s_n^t$. 
Because each configuration corresponds one-to-one to 
the binary number of $n$ digits, the total number of configurations is $ 2^n $. 

\subsection{Master equation}

Let $ p_{\bi{s}}^t $ be the probability of taking configuration $\bi{s}$ 
at time $ t $.  The state of the system at time $t$ is given by state vector
$\underline{p}^t=(p_0^t,p_1^t,\dots,p_{2^n-1}^t)$ 
which evolves in time according to the master equation of the form
$\underline{p}^{t+1}=U\underline{p}^t$, or
\begin{equation}
p^{t+1}_{\bi{s}}=\sum_{\bi{s}'}U_{\bi{s}\bi{s}'}p^t_{\bi{s}'} ,
\label{Master-equation01}
\end{equation}
where $U=\left(U_{\bi{s}\bi{s}'}\right)$ is the $2^n\times 2^n$ transition matrix.
We want to find the $2^n\times 2^n$ transition matrix from time $t$ to $t+1$, $U$. 

\begin{figure}[ht]
\begin{center}
\includegraphics[scale=0.3]{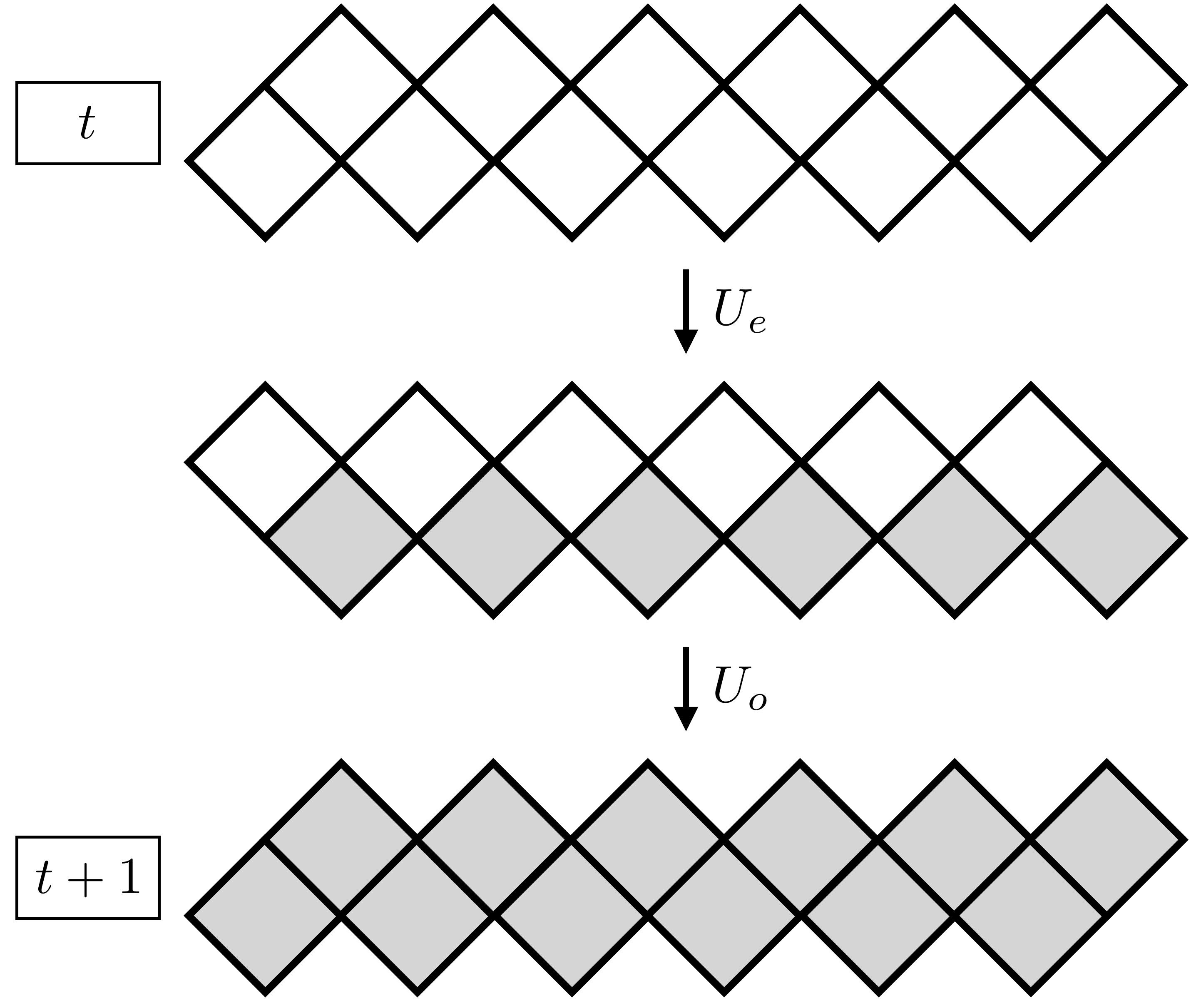}
\end{center}
    \caption{Separation of time evolution. $U_e$ and $U_o$ are transition matrices corresponding to steps (i) and (ii) respectively.}
    \label{figure07}
\end{figure}

As stated in the previous subsection,
the time evolution of RCA54 is divided into two steps: 
(i) values of even-numbered cells except cell $n$ are determined by the rule 
(\ref{CA-rule-even}), and that of cell $n$ by the boundary condition, and 
(ii) values of odd-numbered cells except cell $1$ are determined by the rule 
(\ref{CA-rule-odd}), and that of cell $1$ by the boundary condition. 
Let the configuration at time $t$ be $\bi{s}=s_1\cdots s_n$ and 
that at time $t+1$ be $\bi{u}=u_1 \cdots u_n$.  Then,
step (i) changes configuration $\bi{s}$ to $s_1u_2 s_3 u_4 \cdots u_n $, 
which is transformed into $\bi{u}$ by step (ii) (see Figure \ref{figure07}).
It is noted that in step (i) $u_{2k}$ is determined only from the three 
values $s_{2k-1}$, $s_{2k}$, $s_{2k+1}$ 
and $u_n$ is probabilistically chosen depending on $s_{n-1}$ and $s_n$.
Each transition is represented by small transition matices.

Transition from $s_{2k-1}s_{2k}s_{2k+1}$ to $s_{2k-1}u_{2k}s_{2k+1}$ 
is given by (\ref{CA-rule-even}). 
Interpreting this transition as a 3-digit binary number to a 
3-digit binary number, the local transition matrix is written using the
following $ 8\times8 $ matrix 
\begin{equation}
 P=\left( 
\begin{array}{cccc|cccc}
 1&&& &&&& \\
 &&&1 &&&& \\
 &&1& &&&& \\
 &1&& &&&& \\
 \hline
 &&&& &&1& \\
 &&&& &&&1 \\
 &&&& 1&&& \\
&&&& &1&& \\  
\end{array}\right),
\end{equation}
or tensor
\begin{equation}
 P_{uu'u''|ss's''}=\delta_{us}\delta_{u'\chi(s,s',s'')}\delta_{u''s''}.
\end{equation}
The rows and columns are labeled $ss's''$ 
(in order of $000,001,010,011,100,101,110,111$). 
The same transition matrix is also utilized in step (ii).

The transition matrix at the boundary is obtained as follows. 
The absorption/emission part is expressed by $2\times 2$ matrices 
that acts only on the end cells $1$ or $n$, as
\begin{equation}
 B^L=\left( 
\begin{array}{cccc}
 1-\alpha & \beta \\
 \alpha & 1-\beta
\end{array}\right),\qquad
 B^R=\left( 
\begin{array}{cccc}
 1-\gamma & \delta \\
 \gamma & 1-\delta
\end{array}\right) .
\end{equation}
Next we consider the effect of virtual cell $0$ on the left boundary 
block $s_1u_2$.
When cell $0$ takes value $u_0=0$ with probability $\zeta$, 
the upper left half of the matrix  $P$ acts on $s_1u_2 $. 
On the other hand, if $ u_0=1$ with probability $1-\zeta$, 
the bottom right half of $P$ acts on $s_1u_2$. 
Thus, the transition matrix on the left boundary cells $s_1s_2$ is
\begin{eqnarray}
 \tilde{P}^L&=&\zeta\left( 
\begin{array}{cccc}
 1&&& \\
 &&&1 \\
 &&1& \\
 &1&&
\end{array}\right)
+(1-\zeta)\left( 
\begin{array}{cccc}
 &&1& \\
 &&&1 \\
 1&&& \\
 &1&&
\end{array}\right)\nonumber\\
&=&\left(\begin{array}{cccc}
 \zeta&&1-\zeta& \\
 &&&1 \\
 1-\zeta&&\zeta& \\
 &1&&
\end{array}\right)
.
 \label{boundary-condition-01}
\end{eqnarray}
By a similar argument, it is known that
the effect of the right virtual cell is described by the following matrix: 
\begin{equation}
\tilde{P}^R=\left(
\begin{array}{cccc}
 \eta&1-\eta&& \\
 1-\eta&\eta&& \\
 &&&1 \\
 &&1&
\end{array}\right).
 \label{boundary-condition-02}
\end{equation}

The transition matrices at both boundaries are the compositions 
of $P^L$ and $B^L$ (left), $P^R$ and $B^R$ (right) as follows
\begin{equation}
 P^L=\tilde{P}^L(\mathbf{1}_2\otimes B^L)=\left( 
\begin{array}{cccc}
 q_1^L&&1-q_2^L& \\
 &\alpha&&1-\beta \\
1-q_1^L &&q_2^L& \\
 &1-\alpha&&\beta
\end{array}\right),
\end{equation}
\begin{equation}
 P^R=\tilde{P}^R(B^R\otimes \mathbf{1}_2)=\left( 
\begin{array}{cccc}
 q_1^R&1-q_2^R&& \\
1-q_1^R &q_2^R&& \\
 &&\gamma&1-\delta \\
 &&1-\gamma&\delta
\end{array}\right).
\end{equation}
Here, $\mathbf{1}_2$ is a $2\times 2$ unit matrix, and 
\begin{eqnarray}
 q_1^L&=&\zeta+\alpha-2\zeta\alpha ,\qquad q_2^L= \zeta+\beta-2\zeta\beta ,\\
  q_1^R&=&\eta+\gamma-2\eta\gamma  ,\qquad  q_2^R=\eta+\delta-2\eta\delta.
\end{eqnarray}
Since  
$0\le \alpha, \beta, \gamma, \delta, \zeta, \eta \le1$, these parameters also
satisfy $0 \le q_1^L,\, q_2^L,\, q_1^R,\, q_2^R \le1$.

The transition matrix $U$ can be written as
\begin{equation}
 U=U_oU_e,
\end{equation}
\begin{eqnarray}
 U_e&=& P_{123}P_{345}P_{567}\cdots P_{n-3,n-2,n-1}P^R_{n-1,n}, \\
 U_o&=& P^L_{12}P_{234}P_{456}P_{678}\cdots P_{n-2,n-1,n},
\end{eqnarray}
where
\begin{equation}
 P_{i-1,i,i+1}=\mathbf{1}_{2^{i-2}}\otimes P \otimes \mathbf{1}_{2^{n-i-1}},
\end{equation}
and
\begin{equation}
 P^L_{12}=P_L\otimes \mathbf{1}_{2^{n-2}},\qquad 
 P^R_{n-1,n}=\mathbf{1}_{2^{n-2}}\otimes P^R.
\end{equation}
While $U_e$ is the transition matrix corresponding to step (i), 
$U_o$ is the transition matrix corresponding to step (ii). 
In this way the master equation for the system (\ref{Master-equation01}) is given. 
Note that all components of $U$ are nonnegative and the sum of each column is 1.  
Namely, $U$ is a stochastic matrix.

\subsection{Nonequilibrium steady state}

Although it is difficult to solve the master equation (\ref{Master-equation01}) 
in general, it is possible to solve for the steady state solution exactly.
Here we describe the method of \cite{Prosen16}.
The steady state is the solution of the equation
\begin{equation}
 \underline{p}=U\underline{p}
 \label{NESS-Master-equation00}
\end{equation}
or
\begin{equation}
 p_{\bi{s}}=\sum_{\bi{s}'}U_{\bi{s}\bi{s}'}p_{\bi{s}'} .
 \label{NESS-Master-equation01}
\end{equation}
Theses equations mean that $\underline{p}$ is an eigenvector 
with eigenvalue $1$ of the transition matrix $U$. 
We refer to (\ref{NESS-Master-equation01}) as a NESS equation, and to 
$\underline{p}$ which satisfies (\ref{NESS-Master-equation00}) as a NESS vector.
The existence and uniqueness of the NESS is guaranteed by 
the following theorem proved in \cite{Prosen16}.
\begin{theo}
\label{Theo01}
For all parameters $0<\alpha,\, \beta,\, \gamma,\, \delta, \,\zeta,\, \eta <1$, 
the transition matrix $U$ is irreducible and aperiodic.
\end{theo}
Here, \textit{irreducible} means that there is a certain natural number 
$t_0\in\mathbb{N}$ such that 
$(U^{t_0})_{\bi{s}'\bi{s}}>0$ for any $\bi{s}$ and $\bi{s}'$ and
\textit{aperiodic} means that for any $\bi{s}$, the greatest common 
divisor of all $ t $ such that $(U ^ t)_{\bi{s}\bi{s}}> 0$ is $1$.
We do not prove this theorem here but
according to the Perron-Frobenius's theorem, 
when $U$ is irreducible and aperiodic, 
the maximum eigenvalue of a stochastic matrix is $1$ and 
its eigenspace is one dimension \cite{Gantmacher}.
Thus, the solution of the NESS equation (\ref{NESS-Master-equation01})  is unique. 

We divide the master equation into
\begin{equation}
  p_{\bi{s}}=\sum_{\bi{s}'}(U_o)_{\bi{s}\bi{s}'}p'_{\bi{s}'} \qquad  p'_{\bi{s}}=\sum_{\bi{s}'}(U_e)_{\bi{s}\bi{s}'}p_{\bi{s}'} ,
  \label{devided-NESS-equation}
\end{equation}
and assume the following patch state ansatz (PSA) that the solution can be 
written of the form:
\begin{equation}
 p_{\bi{s}}=L_{s_1s_2s_3}X_{s_2s_3s_4s_5}X_{s_4s_5s_6s_7}\cdots X_{s_{n-4}s_{n-3}s_{n-2}s_{n-1}}R_{s_{n-2}s_{n-1}s_n},
 \label{PSA01}
\end{equation}
\begin{equation}
 p'_{\bi{s}}=L'_{s_1s_2s_3}X'_{s_2s_3s_4s_5}X'_{s_4s_5s_6s_7}\cdots X'_{s_{n-4}s_{n-3}s_{n-2}s_{n-1}}R'_{s_{n-2}s_{n-1}s_n}.
  \label{PSA02}
\end{equation}
Here, $ X, X '$ are rank-4 tensors, and $ L, L', R, R '$ are rank-3 tensors 
of nonnegative components, 
which amount to  $16\times2+8\times 4 = 64 $ unknown variables.

\subsubsection*{Normalisations: }

When the configuration is a vacuum state $\bi{s}=0\cdots 0$, 
(\ref{PSA01}) and (\ref{PSA02}) are
\begin{equation}
 p_{0\cdots0}=L_{000}(X_{0000})^{n/2-2}R_{000},\qquad  p'_{0\cdots0}=L'_{000}(X'_{0000})^{n/2-2}R'_{000}.
\end{equation}
$U_e$ can change value of cell $n$ and $U_o$ value of cell $1$, 
but the rest remains $0$. Hence,
\begin{equation}
L_{000}(X_{0000})^{n/2-2}R_{000}=\sum_{s_n}(U_e)_{0\cdots 0s_n}L'_{000}(X'_{0000})^{n/2-2}R'_{00s_n}.
\end{equation}
Since this is true even if $n$ change to $n+2$, we can see by taking ratio of 
the two equations, 
\begin{equation}
 X_{0000}=X'_{0000}
\end{equation}
Since the degree of freedom of constant multiplication is allowed for the partition function (total sum of probabilities), we can choose $X_{0000}=1$.

\subsubsection*{Gauge symmetry: }

PSA (\ref{PSA01}) and (\ref{PSA02}) have gauge symmetry that makes the equation invariant.
Let $ f_ {ss'}, f '_ {ss'} $ be arbitrary 2-tensors. 
The NESS equation is then invariant under the following gauge transformation:  
\begin{eqnarray}
 X_{ss'tt'}&\mapsto f_{uu'}X_{ss'tt'}f^{-1}_{tt'} ,\nonumber\\
 L_{tss'}&\mapsto  L_{tss'}f^{-1}_{ss'}, \\
 R_{ss't}&\mapsto  f_{ss'}R_{ss't},\nonumber
 \end{eqnarray}
\begin{eqnarray}
 X'_{ss'tt'}&\mapsto f'_{uu'}X'_{ss'tt'}f'^{-1}_{tt'} ,\nonumber\\
 L'_{tss'}&\mapsto  L'_{tss'}f'^{-1}_{ss'} ,\\
 R'_{ss't}&\mapsto  f'_{ss'}R'_{ss't} .\nonumber
\end{eqnarray}
So if we fix gauge to 
\begin{eqnarray}
 f_{ss'}&=&(L_{000},\,X_{0001},\,X_{0010},\,X_{0011}), \\
 f'_{ss'}&=&(R'^{-1}_{000},\,X'_{0001},\,X'_{0010},\,X'_{0011}),
\end{eqnarray}
we can transform the PSA components into
\begin{equation}
 L_{000},\, R'_{000},\,X_{00ss'},\,X'_{00ss'}\mapsto1.
\end{equation}
Note that under this gauge transformation, the diagonal components are invariant, $X_{ss's'}\mapsto f_{ss'}X_{ss'ss'}f^{-1}_{ss'}= X_{ss'ss'}$.

Thus, the number of unknown variables in the equations is reduced 
to $64 - 2 - 2 - 6 = 54$. 
If we successfully determine the 54 unknowns and get a solution for the equations, 
PSA gives the unique solution.

\subsubsection*{Matrix representation of the tensors: }

Here we express the tensors in PSA in matrix form. 
$X_{ss' tt '},\,X'_{ss'tt'}$ can be represented by $4\times4$ matrices whose 
rows are labeled by $ss' \,$(in order of $00, 01, 10, 11$) and whose columns labeled by $tt'$. 
Then, let unknown components of the matrices  be 
\begin{equation}
 X=\left(\begin{array}{cccc}
 1&1&1&1 \\
y_1 &y_2&y_3&y_4 \\
y_5&y_6&y_7&y_8 \\
y_9 &y_{10}&y_{11}&y_{12}
\end{array}\right),
\end{equation}
\begin{equation}
 X'=\left(\begin{array}{cccc}
 1&1&1&1 \\
y'_1 &y'_2&y'_3&y'_4 \\
y'_5 &y'_6&y'_7&y'_8 \\
y'_9 &y'_{10}&y'_{11}&y'_{12}
\end{array}\right).
\end{equation}
Similarly $L_{tss'},\,L'_{tss'}$ and $R_{ss't},\,R'_{ss't}$ can be represented by 
using $2\times4,\,4\times2$ matrices respectively, 
\begin{equation}
 L=\left(\begin{array}{cccc}
 1&l_2&l_3&l_4 \\
l_5&l_6&l_7&l_8
\end{array}\right),
\qquad
L'=\left(\begin{array}{cccc}
 l_1'&l_2'&l_3'&l_4' \\
l_5'&l_6'&l_7'&l_8'
\end{array}\right),
\end{equation}
\begin{equation}
 R=\left(\begin{array}{cc}
 r_1&r_2 \\
 r_3&r_4 \\
 r_5&r_6 \\
 r_7&r_8
\end{array}\right),
\qquad
 R'=\left(\begin{array}{cc}
 1&r_2' \\
 r_3'&r_4' \\
 r_5'&r_6' \\
 r_7'&r_8'
\end{array}\right).
\end{equation}

\subsubsection*{Reduced NESS equations: }

The 54 unknowns can be determined by examining specific components of the NESS vectors. 
First, we focus on the following components: 
(I) all indices except $s_1,\, s_2,\, s_3$ are $0$, 
(II) all indices except $s_{n-2},s_{n-1},s_n$ are $0$, 
(III) all indices except $s_{2k+2},s_{2k+3},s_{2k+4},s_{2k+5},\,(0<k<n/2-3)$ are $0$.
In each case, calculating the NESS equation (\ref{NESS-Master-equation01}) directly, 
we obtain
\begin{equation}
 L'_{s_1s_2s_3}X'_{s_2s_300}=\sum_{t_n}(P^R)_{000t_n}R_{00t_n}L_{s_1\chi(s_1s_2s_3)s_3}X_{\chi(s_1s_2s_3)s_3s_30}X_{s_3000},
\end{equation}
\begin{equation}
 L_{s_1s_2s_3}X_{s_2s_300}R_{000}=\sum_{t_1}(P^L)_{s_1s_2t_1s_2}L'_{t_1s_2\chi(s_2s_30)}X'_{s_2\chi(s_2s_30)00},
\end{equation}
\begin{equation}
 L'_{000}R'_{s_{n-2}s_{n-1}s_n}=\sum_{t_n}(P^R)_{s_{n-1}s_ns_{n-1}t_n},
\end{equation}
\begin{equation}
 R_{s_{n-2}s_{n-1}s_n}=\sum_{t_1}(P^L)_{00t_10}L'_{t_100}X'_{0s_{n-2}s_{n-2}\chi(s_{n-2}s_{n-1}s_n)}R'_{s_{n-2}\chi(s_{n-2}s_{n-1}s_n)s_n},
\end{equation}
\begin{eqnarray}
&L'_{000}X'_{s_2s_3s_4s_5}X'_{s_4s_500} \nonumber\\
&=\sum_{t_n}(P^R)_{000t_n}R_{00t_n}X_{\chi(0s_2s_3)s_3\chi(s_3s_4s_5)s_5}X_{\chi(s_3s_4s_5)s_5s_50}X_{s_5000}  ,
\label{reduced-NESS-equation05}
\end{eqnarray}
\begin{eqnarray}
&X_{s_2s_3s_4s_5}X_{s_4s_500}R_{000} \nonumber\\
&=\sum_{t_1}(P^L)_{00t_10}L'_{t_100}X_{0s_2s_2\chi(s_2s_3s_4)}X_{s_2\chi(s_2s_3s_4)s_4\chi(s_4s_50)}X_{s_4\chi(s_4s_50)00}  .
\label{reduced-NESS-equation06}
\end{eqnarray}
In (\ref{reduced-NESS-equation05}) and (\ref{reduced-NESS-equation06}), 
we have replaced $s_{2k+2},s_{2k+3},s_{2k+4},s_{2k+5}$ by $s_2s_3s_4s_5$ for simplicity
of presentation.
Solving these equations, we have
\[
 X=\left(\begin{array}{cccc}
 1&1&1&1 \\
y_1 &y_2&y_3&y_4 \\
y_1&y_4/y_3&y_2&y_4/y_3 \\
y_1/y_4 &1/y_3&1&y_2
\end{array}\right),
\]

\[
 X'=\left(\begin{array}{cccc}
 1&1&1&1 \\
y_1 &y_2&1/y_4&1/y_3 \\
y_1 &y_4/y_3&y_2&y_4/y_3 \\
y_1y_3 &y_4&1&y'_2
\end{array}\right),
\]
\[
 L=\left(\begin{array}{cccc}
 1&l_2&l_3&l_4 \\
l_5&l_6&l_7&l_8
\end{array}\right),
\qquad
L'=\left(\begin{array}{cccc}
 l_1'&l_1'l_4&l_1'l_3&l_1'l_2 \\
l_1'l_7y_1&l_1l_8'&l_1'l_5/y_1&l_1'l_6
\end{array}\right),
\]
\[
 R=\left(\begin{array}{cc}
 r_1&r_2 \\
 r_3&r_4 \\
 r_5&r_6 \\
 r_7&r_8
\end{array}\right),
\qquad
 R'=\left(\begin{array}{cc}
 1&r_4/r_1 \\
r_3/r_1 & r_2/r_1 \\
y_1y_3 r_7/r_1 & y_1y_3 r_8/r_1 \\
r_5/(y_3r_1) & r_6/(y_3 r_1)
\end{array}\right).
\]
There are still remained $4+7+1+8=20$ undetermined variables in PSA, which are related by
\begin{eqnarray}
  l_1'&=q_1^R r_1+(1-q_2^R)r_2 ,\nonumber\\
 l_1'r_2&=r_1[(1-\gamma)r_7+\delta r_8] ,\nonumber\\
 l_1'r_3&=r_1[\gamma r_7+(1-\delta)r_8] ,\nonumber\\
 l_1'r_4&=r_1[(1-q_1^R)r_1+q_2^Rr_2] ,\nonumber\\
 l_1'r_5&=r_1/y_3\cdot[\gamma r_3+(1-\delta)r_4] ,\nonumber\\
 l_1'r_6&=r_1/y_3\cdot[(1-\gamma)r_3+\delta r_4], \nonumber\\
 l_1'r_7&=r_1/y_4\cdot[q_1^Rr_5+(1-q_2^R)r_6], \nonumber\\
 l_1'r_8&=r_1/y_4\cdot[(1-q_1^R)r_5+q_2^Rr_6] ,
\label{NESS-equation-Rsector}
\end{eqnarray}
\begin{eqnarray}
 r_1&=l_1'[q_1^L+(1-q_2^L)l_7y_1] ,\nonumber\\
 r_1l_2&=l_1'[q_1^Ll_4+(1-q_2^L)l_8] ,\nonumber\\
 r_1l_3&=l_1'y_3[\alpha l_2+(1-\beta)l_6], \nonumber\\
 r_1l_4&=l_1'y_4/y_1\cdot[\alpha l_3y_1+(1-\beta)l_5],\nonumber \\
 r_1l_5&=l_1'[1-q_1^L+q_2^Ll_7y_1] ,\nonumber\\
 r_1l_6&=l_1'[(1-q_1^L)l_4+q_2^Ll_8] ,\nonumber\\
 r_1l_7&=l_1'y_3[(1-\alpha)l_2+\beta l_6] ,\nonumber\\
 r_1l_8&=l_1'y_4/y_1\cdot[(1-\alpha)l_3y_1+\beta l_5].
\label{NESS-equation-Lsector}
\end{eqnarray}
However, since all unknowns cannot be determined by these equations alone, 
other component equations are required.

Next, we focus on the following components: 
(IV) all indices except $s_{2k+2},s_{2k+3},s_{2k+4},s_{2k+5},\,(0<k<n/2-3)$ are $1$, 
(V) all indices except $s_2,s_3,s_4,s_5,\,(k=0)$ are $0$, 
(VI) all indices except $s_{n-4},s_{n-3},s_{n-2},s_{n-1},\,(k=n/2-3)$ are $0$. 
From these components, we obtain the following equations,
\begin{eqnarray}
  y_2&=y_1 ,\nonumber\\
  y_4&=y_1y_3 ,\nonumber\\
  l_1'y_1&=q_1^Rr_5+(1-q_2^R)r_6 ,\nonumber\\
  l_1'r_5'&=y_1[q_1^Rr_1+(1-q_2^R)r_2] ,\nonumber\\
  r_1&=l_1'[q_1^Ll_4+(1-q_2^L)l_8] ,\nonumber\\
  r_1l_2&=l_1'[q_1^L+(1-q_2^L)^Ll_7y_1].
\label{NESS-equation-Extra}
\end{eqnarray}
Thus, we can represent $y_2$ and $y_4$ using $y_1$ and $y_3$, 
and moreover find relations $l_2 = 1$ and $r_7=r_1/y_3$.
The remaining unknowns are $y_1$, $y_3$, $l_3$,\dots, $l_8$, $r_1$, \dots,$r_8$, 
and $l_1'$.
Because all components of $R$ and $L'$ include $l_1'$ as a factor, and the PSA also does,
we can set $l_1'=1$ using the freedom of multiplying a constant to the eigenvector.

\subsubsection*{Exact solutions of the reduced NESS equations: }

It is still difficult to solve 
(\ref{NESS-equation-Rsector}) and (\ref{NESS-equation-Lsector}).
However, there are cases where a special relation holds between parameters,
the equations become simpler and an exact solution is obtained.
In the previous study \cite{Prosen16}, the exact solution for $\zeta=\eta =1/2$ is given.
Here we will show that an exact solution is obtained for arbitrary $\zeta$ and $\eta$ 
if the conditions
\begin{equation}
  \alpha=1-\beta,\quad \gamma=1-\delta
\label{eq-solvability}
\end{equation}
is satisfied.
This condition means that when updating the boundary cells, 
the probability of $s_1=1$ is $\alpha$ and that of $s_n=1$ is
$\gamma$ \textit{irrespective of the previous values}.
Both of the two cases are united into $q_1^L=1-q_2^L$ and $q_1^R=1-q_2^R$, 
and the linear combinations of $r_i$ and $l_i$ in the right hand side 
of (\ref{NESS-equation-Rsector}) and (\ref{NESS-equation-Lsector}) become simple.
Especially in the former choice, $q_1^L=q_1^R=1/2$.

When  
$q_1^L=1-q_2^L$ and $q_1^R=1-q_2^R$  as the NESS equations are
\begin{eqnarray}
  1&=q_1^R( r_1+r_2), \nonumber\\
 r_2&=r_1(1-\gamma)(r_7+ r_8) ,\nonumber\\
 r_3&=r_1\gamma (r_7+r_8),\nonumber \\
 r_4&=r_1(1-q_1^R)(r_1+r_2),\nonumber \\
 r_5&=r_1/y_3\cdot\gamma (r_3+r_4) ,\nonumber\\
 r_6&=r_1/y_3\cdot(1-\gamma)(r_3+ r_4) ,\nonumber\\
 r_7&=r_1/(y_1y_3)\cdot q_1^R(r_5+r_6) ,\nonumber\\
 r_8&=r_1/(y_1y_3)\cdot(1-q_1^R)(r_5+r_6) ,
\end{eqnarray}
\begin{eqnarray}
 r_1&=q_1^L(1+l_7y_1),\nonumber \\
 r_1&=q_1^L(l_4+l_8) ,\nonumber\\
 r_1l_3&=y_3\alpha (1+l_6) ,\nonumber\\
 r_1l_4&=y_3\alpha (l_3y_1+l_5) ,\nonumber\\
 r_1l_5&=(1-q_1^L)(1+l_7y_1),\nonumber \\
 r_1l_6&=(1-q_1^L)(l_4+l_8) ,\nonumber\\
 r_1l_7&=y_3(1-\alpha)(1+ l_6) ,\nonumber\\
 r_1l_8&=y_3(1-\alpha)(l_3y_1+ l_5),
\end{eqnarray}
\begin{equation}
 r_7=r_1/y_3.
\end{equation}
Then we arrive at the same form of transfer matrices as (43) in \cite{Prosen16}
\begin{equation}
 X=\left(
\begin{array}[tb]{cccc}
 1& 1& 1& 1\\
\xi\omega & \xi\omega & \xi^{-1} & \omega \\
\xi\omega & \xi\omega & \xi\omega & \xi\omega \\
\xi & \xi & 1 & \xi\omega
\end{array}
\right),\quad
 X'=\left(
\begin{array}[tb]{cccc}
 1& 1& 1& 1\\
\xi\omega & \xi\omega & \omega^{-1} & \xi \\
\xi\omega & \xi\omega & \xi\omega & \xi\omega \\
\omega & \omega & 1 & \xi\omega
\end{array}
\right)
\label{Transfer-matrices-obtained}
\end{equation}
where
\begin{equation}
 \xi:=\frac1{y_3}= \frac{(1-\gamma)\alpha+(1-\alpha)q_1^R}{[(1-\alpha)\gamma+(1-\gamma)q_1^L]^2}
 \Bigl(\alpha+(1-\alpha)q_1^R-q_1^Lq_1^R\Bigr),
 \label{exact-NESS54-sol-xi}
\end{equation}
and
\begin{equation}
 \omega:=y_1y_3=\frac{(1-\alpha)\gamma+(1-\gamma)q_1^L}{[(1-\gamma)\alpha+(1-\alpha)q_1^R]^2}
 \Bigl(\gamma+(1-\gamma)q_1^L-q_1^Lq_1^R\Bigr),
  \label{exact-NESS54-sol-omega}
\end{equation}
Matrices $L$, $L'$, $R$, and $R'$ are written with the following quantities.
\begin{eqnarray}
 r_1&=\frac{(1-\alpha)\gamma+(1-\gamma)q_1^L}{(1-\gamma)\alpha+(1-\alpha)q_1^R} ,\nonumber\\
 r_2&=\frac{1-\gamma}{q_1^R[(1-\gamma)\alpha+(1-\alpha)q_1^R]}\Bigl(\alpha+(1-\alpha)q_1^R-q_1^Lq_1^R\Bigr) ,\nonumber\\
 r_3&=   \frac{\gamma}{q_1^R[(1-\gamma)\alpha+(1-\alpha)q_1^R]}\Bigl(\alpha+(1-\alpha)q_1^R-q_1^Lq_1^R\Bigr) ,\nonumber\\
 r_4&= \left(\frac{1}{q_1^R}-1\right)\frac{(1-\alpha)\gamma+(1-\gamma)q_1^L}{(1-\gamma)\alpha+(1-\alpha)q_1^R} ,\nonumber\\
 r_5&= \frac{\gamma}{q_1^R}\frac{[\alpha+(1-\alpha)q_1^R-q_1^Lq_1^R][\gamma+(1-\gamma)q_1^L-q_1^Lq_1^R]}{[(1-\alpha)\gamma+(1-\gamma)q_1^L][(1-\gamma)\alpha+(1-\alpha)q_1^R]} ,\nonumber\\
 r_6&= \frac{1-\gamma}{q_1^R}\frac{[\alpha+(1-\alpha)q_1^R-q_1^Lq_1^R][\gamma+(1-\gamma)q_1^L-q_1^Lq_1^R]}{[(1-\alpha)\gamma+(1-\gamma)q_1^L][(1-\gamma)\alpha+(1-\alpha)q_1^R]} ,\nonumber\\
 r_7&= \frac{\alpha+(1-\alpha)q_1^R-q_1^Lq_1^R}{(1-\alpha)\gamma+(1-\gamma)q_1^L} ,\nonumber\\
 r_8&=\left(\frac{1}{q_1^R}-1\right) \frac{\alpha+(1-\alpha)q_1^R-q_1^Lq_1^R}{(1-\alpha)\gamma+(1-\gamma)q_1^L} ,
 \label{exact-NESS54-sol-R}
\end{eqnarray}
\begin{eqnarray}
 l_3&=  \frac{\alpha}{q_1^L}\frac{(1-\alpha)\gamma+(1-\gamma)q_1^L}{\alpha+(1-\alpha)q_1^R-q_1^Lq_1^R} ,\nonumber\\
 l_4&=  \frac{\alpha}{q_1^L}\frac{(1-\alpha)\gamma+(1-\gamma)q_1^L}{(1-\gamma)\alpha+(1-\alpha)q_1^R} ,\nonumber\\
 l_5&=l_6=\frac{1}{q_1^L}-1,\nonumber \\
 l_7&=  \frac{1-\alpha}{q_1^L}\frac{(1-\alpha)\gamma+(1-\gamma)q_1^L}{\alpha+(1-\alpha)q_1^R-q_1^Lq_1^R} ,\nonumber\\
 l_8&=  \frac{1-\alpha}{q_1^L}\frac{(1-\alpha)\gamma+(1-\gamma)q_1^L}{(1-\gamma)\alpha+(1-\alpha)q_1^R} .
 \label{exact-NESS54-sol-L}
\end{eqnarray}

\subsection{Physical quantities in the NESS}
\subsubsection*{Partition function: }

Since the NESS vectors $p_{\bi{s}},\,p_{\bi{s}}'$ are obtained, we can calculate the expectation value of physical quantities with these vectors.
Partition functions of the system are
\begin{equation}
 Z_n:=\sum_{\bi{s}}p_{\bi{s}},\qquad Z'_n=\sum_{\bi{s}}p'_{\bi{s}}.
\end{equation}
Substituting the PSA (\ref{PSA01}) and (\ref{PSA02}) gives the following matrix product forms:
\begin{equation}
 Z_n=\bi{l}X^{n/2-2}\bi{r} ,\qquad Z'_n=\bi{l}'X'^{n/2-2}\bi{r}' .
\end{equation}
Here we introduced a four component vectors
\begin{eqnarray}
 \bi{l}:=(L_{0ss'}+L_{1ss'}), \qquad  \bi{l}':=(L'_{0ss'}+L'_{1ss'}),\nonumber\\
 \bi{r}:=(R_{ss'0}+R_{ss'1}) ,\qquad  \bi{r}':=(R'_{ss'0}+R'_{ss'1}).
\end{eqnarray}
In the case of exact solution (\ref{exact-NESS54-sol-xi})-(\ref{exact-NESS54-sol-L}), 
\begin{eqnarray}
  \bi{l}X=\tau_1 \bi{l}\qquad X\bi{r}=\tau_1\bi{r},\nonumber \\
  \bi{l}'X'=\tau_1 \bi{l}'\qquad X'\bi{r}'=\tau_1\bi{r}'   ,
\label{eigenvector-NESS-exact}
\end{eqnarray}
\begin{equation}
 \tau_1=\frac{[1-(1-2\alpha)(1-2\gamma)(1-\zeta)(1-\eta)]^2}{[(1-\gamma)\alpha+(1-\alpha)q_1^R][(1-\alpha)\gamma+(1-\gamma)q_1^L]}
\end{equation}
are satisfied.
That is, $\bi{l}$ and $\bi{r}$ are eigenvectors of $X$ belonging to the eigenvalue $\tau_1$.
By using this fact repeatedly,
\begin{equation}
 Z_n=\tau_1^{n/2-2}\bi{l}\cdot\bi{r} ,   \qquad Z'_n=\tau_1^{n/2-2}\bi{l}'\cdot\bi{r}'   .
\end{equation}
From direct calculation, we obtain
\begin{equation}
 \bi{l}\cdot\bi{r}  = \bi{l}'\cdot\bi{r}'   =\frac{\alpha\gamma-2\alpha-2\gamma-2(1-\alpha)q_1^R-2(1-\gamma)q_1^L+3q_1^Lq_1^R}{q_1^Lq_1^R[(1-\gamma)\alpha-(1-\alpha)q_1^R]}.
\end{equation}
Hence, $Z_n=Z'_n$.
In the following, we calculate expectation values with $Z_n$.

\subsubsection*{Density: }

We define the density of $1$ at cell $i$ as 
\begin{equation}
 \rho_i:=\frac{1}{Z_n}\sum_{\bi{s}}s_ip_{\bi{s}}
 \label{RCA54-density}
\end{equation}
First consider the case $i=2k, \,(k=1,\cdots,\, n/2-1)$. 
Substituting the PSA (\ref{PSA01}), it becomes
\[
\rho_{2k}=\frac{1}{Z_n}\sum_{s_{2k},s_{2k+1}}\Bigl(\bi{l}X^{k-1}\Bigr)_{s_{2k},s_{2k+1}}
s_{2k}\left(X^{n/2-k-1}\bi{r}\right)_{s_{2k},s_{2k+1}}.
\]
Because $X$ is replaced with $\tau_1$ by acting on $\bi{l}$ or $\bi{r}$ 
as in (\ref{eigenvector-NESS-exact}), we have
\[
\rho_{2k}=\frac{\tau_1^{n/2-2}}{Z_n}\sum_{s_{2k},s_{2k+1}} l_{s_{2k}s_{2k+1}}
s_{2k}r_{s_{2k}s_{2k+1}}
\]
Substituting the exact solution, we arrive at 
\begin{equation}
 \rho_{2k}=\frac{-\alpha-\gamma-(1-\alpha)q_1^R-(1-\gamma)q_1^L+2q_1^Lq_1^R}{\alpha\gamma-2\alpha-2\gamma-2(1-\alpha)q_1^R-2(1-\gamma)q_1^L+3q_1^Lq_1^R}.
\end{equation}
Similarly, we can derive that $\rho_{2k+1}=\rho_{2k}$ for $k=1,\dots,n/2-1$.
Thus the density of 1 is uniform in $2\le i\le n-1$ and its value is denoted by $\rho$ 
in the following.

Finally, unlike bulk, the densities at boundary $i=1, n$ are calculated by the formulas
\begin{eqnarray}
 \rho_1&=\frac{\tau_1^{n/2-2}}{Z_n}\sum_{s_1,s_2,s_3} L_{s_1s_2s_3}s_1r_{s_2s_3} ,\nonumber\\
 \rho_n&=\frac{\tau_1^{n/2-2}}{Z_n}\sum_{s_{n-2},s_{n-1},s_n} l_{s_{n-2}s_{n-1}}s_n
R_{s_{n-2}s_{n-1}s_n}
\end{eqnarray}
which produces lengthy equations we do not show here.
Those values are different from the bulk density.

Let us examine the range of values for the bulk density. 
For the emission limit $\alpha,\, \gamma \to1 $, 
the density takes the maximum value $2/3$ regardless of $\zeta,\,\eta$. 
On the other hand, for the absorption limit $\alpha,\,\gamma \to0$, it become
\[
\rho=\frac{2-x}{3-2x},\qquad x=\frac{1}{\zeta}+\frac{1}{\eta}.
\]
In the range of $0<\zeta,\,\eta<1$, $x$ is in $2<x<\infty$ and 
the density is in $0<\rho<1/2$. 
Thus, the density can take a value in the range $0<\rho <2/3$. 
The fact that the maximum value of the density is $2/3$ instead of $1$ is directly 
understood from the rule (\ref{RCA54-def}).  
The rule allows $s_{2k}^t=s_{2k}^{t+1}=1$ only in the case $s_{2k-1}^t=s_{2k+1}^t=0$,
which inevitably leads to $s_{2k-1}^{t+1}=s_{2k+1}^{t+1}=1$ 
because $\chi(*,0,1)=\chi(1,0,*)=1$. 
Then, $s_{2k}^{t+2}=\chi(s_{2k-1}^{t+1},s_{2k}^{t+1},s_{2k+1}^{t+1})=\chi(1,1,1)=0$.
Thus any cell cannot have 1 three times in a row.
Thus the density of 1 is bounded above by $2/3$. 
It is interesting that our solution covers all possible values of the density, whereas 
the bulk steady-state density can only take values in interval $(2/5, 2/3)$ for 
the solution in \cite{Prosen16}

\subsubsection*{Current: }

As in \cite{Bobenko93,Prosen16}, the particle current is defined as the expectation value 
of the density of right-movers minus that of left-movers $J=J_R-J_L$, where
\begin{eqnarray}
 J_R&:=\frac{1}{Z_n}\sum_{\bi{s}}s_{2k}s_{2k+1}p_{\bi{s}} ,\nonumber\\
 J_L&:=\frac{1}{Z_n}\sum_{\bi{s}}s_{2k+1}s_{2k+2}p_{\bi{s}}.
\label{RCA54-current}
\end{eqnarray}
Substituting the PSA(\ref{PSA01}) and using  (\ref{eigenvector-NESS-exact}), 
$J_R$ is obtained as
\begin{eqnarray}
J_R&=&\frac{\tau_1^{n/2-2}}{Z_n}\sum_{s_{2k},s_{2k+1}} l_{s_{2k}s_{2k+1}}s_{2k}s_{2k+1}
r_{s_{2k}s_{2k+1}}\nonumber\\
&=&
\frac{-\alpha-(1-\alpha)q_1^R+q_1^Lq_1^R}
{\alpha\gamma-2\alpha-2\gamma-2(1-\alpha)q_1^R-2(1-\gamma)q_1^L+3q_1^Lq_1^R}.
\end{eqnarray}
Quite similarly, $J_L $ is obtained as
\begin{equation}
J_L=\frac{-\gamma-(1-\gamma)q_1^R+q_1^Lq_1^R}
{\alpha\gamma-2\alpha-2\gamma-2(1-\alpha)q_1^R-2(1-\gamma)q_1^L+3q_1^Lq_1^R}.
\end{equation}
Thus, we arrive at
\begin{equation}
 J=\frac{(\gamma-\alpha)(2-\alpha-\gamma)+\eta(1-\gamma)(1-2\gamma)-\zeta(1-\alpha)(1-2\alpha)}
{\alpha\gamma-2\alpha-2\gamma-2(1-\alpha)q_1^R-2(1-\gamma)q_1^L+3q_1^Lq_1^R}.
\end{equation}
In the case where $\eta=\zeta$ and difference between $\alpha$ and $\gamma$ is
small, the current is approximately proportional to $\gamma-\alpha$.  
 
We notice another interesting relation
\begin{equation}
J_L+J_R=\rho. 
\label{JLJRrho}
\end{equation}
This is also explained in the same manner as for the maximum value of the bulk density.
Assume $(s_{2k-1}^t,s_{2k}^t,s_{2k+1}^t)=(0,1,0)$. 
Then, $s_{2k}^{t+1}=\chi(s_{2k-1}^t,s_{2k}^t,s_{2k+1}^t)=1$, and 
because $\chi(*,0,1)=\chi(1,0,*)=1$, 
$s_{2k-1}^{t+1}=\chi(s_{2k-2}^{t+1},s_{2k-1}^{t},s_{2k}^{t+1})=1$ and 
$s_{2k+1}^{t+1}=\chi(s_{2k}^{t+1},s_{2k+1}^t,s_{2k+2}^{t+1})=1$, namely
we obtain $(s_{2k-1}^{t+1},s_{2k}^{t+1},s_{2k+1}^{t+1})=(1,1,1)$.
The time-reversal symmetry of the rule (\ref{time-reversal-RCA54}) ensures that the
inverse is true; if $(s_{2k-1}^{t+1},s_{2k}^{t+1},s_{2k+1}^{t+1})=(1,1,1)$, we have
$(s_{2k-1}^t,s_{2k}^t,s_{2k+1}^t)=(0,1,0)$. Similarly, we can derive that
$(s_{2k}^{t+1},s_{2k+1}^{t+1},s_{2k+2}^{t+1})=(0,1,0)$ if and only if 
$(s_{2k}^t,s_{2k+1}^t,s_{2k+2}^t)=(1,1,1)$.  These properties leads to that
the marginal distribution $p(s_{i-1},s_{i},s_{i+1})$ in a stationary state must satisfy
$p(0,1,0)=p(1,1,1)$ for $2\le i\le N-2$.  Using the marginal distribution, 
we can write $J_R=p(1,1,0)+p(1,1,1)$,
$J_L=p(0,1,1)+p(1,1,1)$ and $\rho_i=p(0,1,0)+p(1,1,0)+p(0,1,1)+p(1,1,1)$.
Thus the relation (\ref{JLJRrho}) is derived from $p(0,1,0)=p(1,1,1)$.
This relation must be satisfied in any stationary states.

\section{Temperature driven RCA 54}\label{chapter-heatbath}

\subsection{The relation between RCA54 and ERCA 250R}

\begin{figure}[ht]
\begin{center}
\includegraphics[scale=0.3]{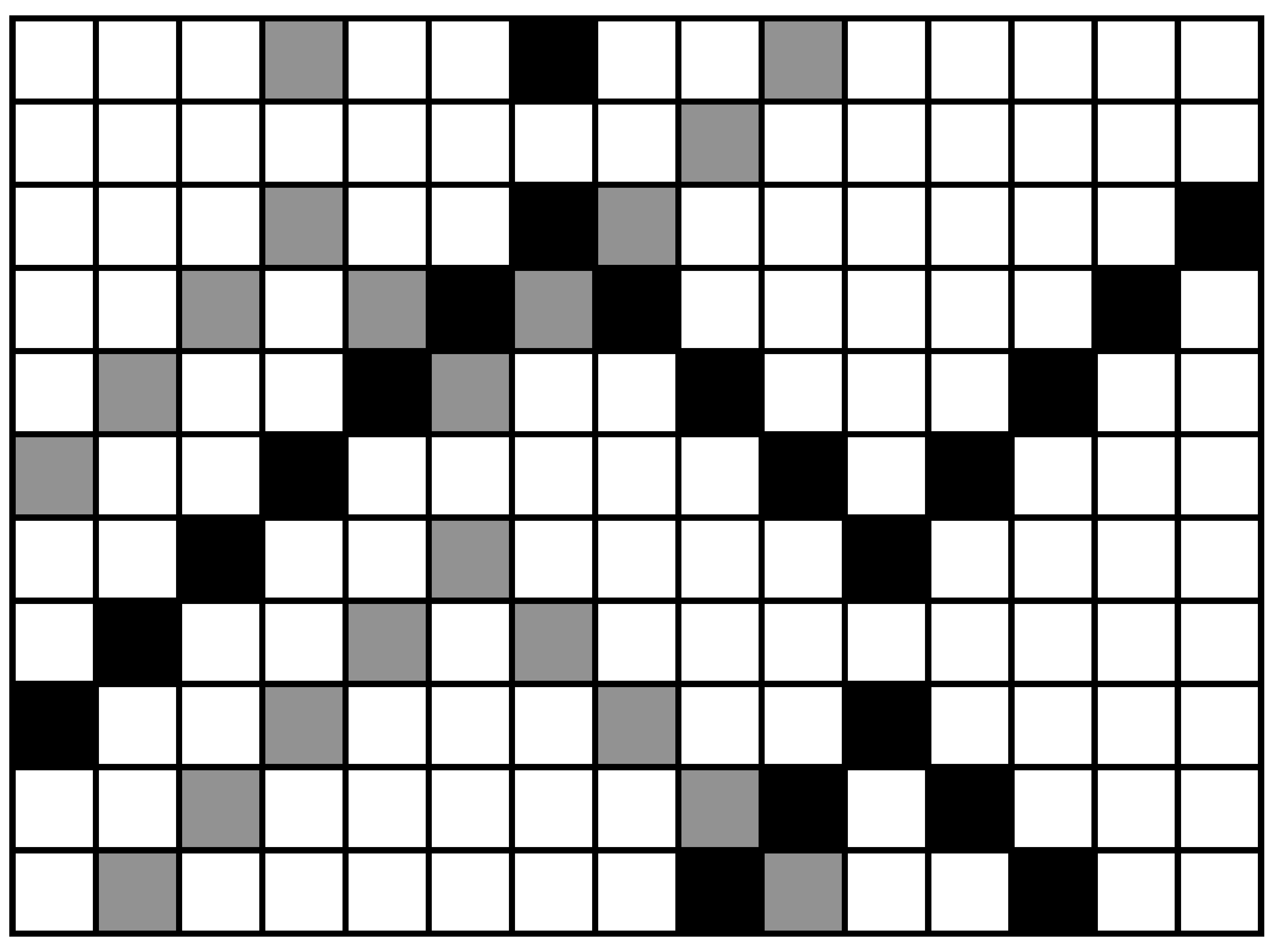}
\end{center}
    \caption{Time evolution of $s_i$ according to ERCA rule 250R. 
    Each column represents the configuration at each time and the time flows downward. White represents the value $0$, black and gray represent the value $1$. It can be seen that blacks collide and shift by two steps, and grays behave similarly, but black and gray pass through.}
    \label{figure09}
\end{figure}

Before introducing the boundary condition, we discuss the relation between 
Bobenko's RCA54 and ERCA 250R.  As mentioned in Introduction, ERCA 250R
is given by (\ref{ERCA}) with $f(0,0,0)=f(0,1,0)=0$ and $f(x,y,z)=1$ for
all other combinations of $(x,y,z)$.  It is also represented as
\begin{equation}
 x_i^{\tau+1}=x_{i-1}^\tau \oplus x_{i+1}^{\tau} \oplus x_{i-1}^\tau 
x_{i+1}^\tau                   \oplus x_i^{\tau-1},
\end{equation}
where we have changed $t$ to $\tau$.
It should be noted that the right hand side of this equation does not depend
on $x_{i}^{\tau}$.  This property leads to the fact that spatiotemporal evolution
$\left\{x_i^\tau\right\}$ with $i+\tau=\mathrm{even}$ and that with 
$i+\tau=\mathrm{odd}$ are independent of each other.
In the former set, if we write $s_{2k}^{t}=x_{2k}^{2t}$ and 
$s_{2k+1}^{t}=x_{2k+1}^{2t+1}$, $s_{2k}^{t+1}$ is
determined by (\ref{CA-rule-even}) and $s_{2k+1}^{t+1}$ is
determined by (\ref{CA-rule-odd}). 
The set $\{x_i^\tau\}$ with $i+\tau=\mathrm{odd}$ is similarly regarded 
as the evolution of RCA54.
That is, ERCA rule 250R is decomposed into two independent RCA54.
It is illustrated in Figure \ref{figure09}.

\subsection{Heat-bath boundary conditions}

We introduce here a heat-bath boundary condition to RCA54 in the same manner as
is done in \cite{Takesue90} for ERCA.  To do so, we have to identify an additive
conserved quantity for RCA54.   It is shown in \cite{Hattori90} that ERCA rule 250R
has several such conserved quantities.  We adapt the result to RCA54 and obtain that
\begin{equation}
 E_{i}^t:=(-1)^{i+1}|s^t_i-s^t_{i+1}|
=\left\{ 
\begin{array}{cl}
 (-1)^{i+1} & s_i\neq s_{i+1}\\
 0 & s_i=s_{i+1}
\end{array}\right. .
\label{energy}
\end{equation}
is a conserved density, which is interpreted as energy in the following.
Note that the energy is not carried by particles but lies between cells.
The conservation can be verified directly from the equality of $\chi$ 
\begin{equation}
 |x-y|-|y-z|=|\chi(x,y,z)-z|-|x-\chi(x,y,z)|,
\end{equation}
which means that the sum of energy is conserved for each diamond-shaped
plaquette.
The conservation law is also written in the form of  the equation of continuity
\begin{equation}
 E^{t+1}_i-E^{t}_i=-\left(S_{i+1}^t-S^t_i\right).
 \label{eq-continuity}
\end{equation}
with energy current
 \begin{equation}
 S_i^t:=1-s^t_{i}-s^{t+1}_i.
\label{eq-energyflux}
\end{equation}
Though the constant in the right-hand side can be chosen arbitrarily, 
we set it to unity for later convenience.

\begin{figure}[ht]
\begin{center}
\includegraphics[scale=0.25]{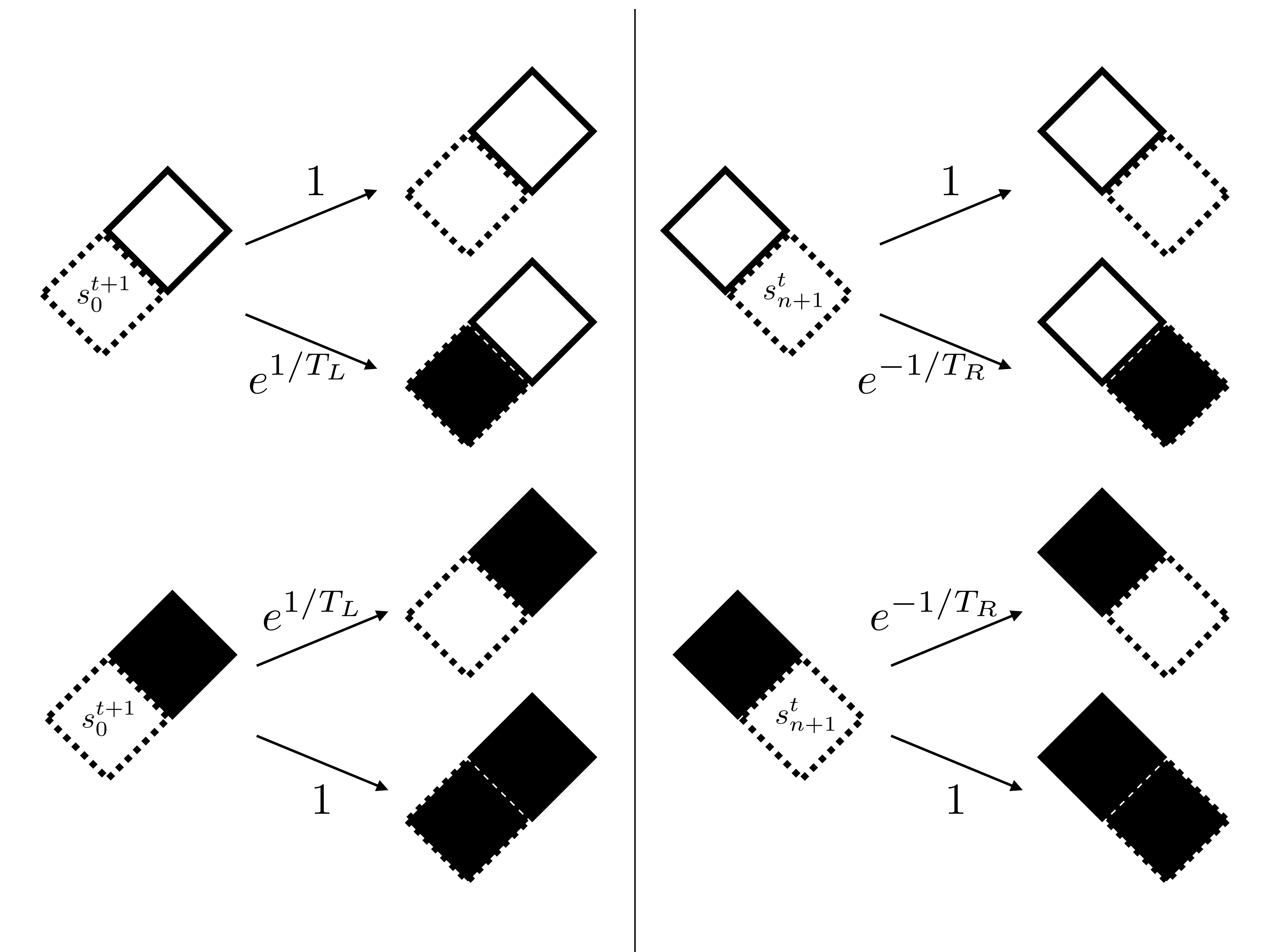}
\end{center}
    \caption{Boundary conditions assumed that thermal bath is in contact. 
    The weights of the values taken by the virtual cells are determined according to the values of the cells $1,\,n $.}
    \label{figure08}
\end{figure}

Next, we devise the heat-bath boundary condition by assigning Gibbs weights for the 
values of the virtual cells $0$ and $n + 1$:
\begin{equation}
 P_{s_0}=\frac{1}{N_L}e^{-E_{0}/T_L} ,\qquad  P_{s_{n+1}}=\frac{1}{N_R}e^{-E_{n}/T_R}.
\end{equation}
Here, $T_L$ and $T_R$ mean the ``temperature" of the thermal baths in contact
at the left and right boundaries, respectively, and 
$N_L=1+e^{1/T_L}$ and $N_R=1+e^{-1/T_R}$ are normalization factors. 
Note that because the energy is bounded, $ T_L$ and $T_R$ can take negative values.
In this boundary condition, $\tilde{P}^L$ in (\ref{boundary-condition-01}) 
and $\tilde{P}^R$ in (\ref{boundary-condition-02}) are replaced by
\begin{eqnarray}
  \tilde{P}^L&=\left(\begin{array}{cccc}
 1/N_L&&1/N_L& \\
 &&&1 \\
 e^{1/T_L}/N_L&&e^{1/T_L}/N_L& \\
 &1&&
\end{array}\right),\nonumber\\
\tilde{P}^R&=\left(\begin{array}{cccc}
 1/N_R&1/N_R&& \\
 e^{-1/T_R}/N_R&e^{-1/T_R}/N_R&& \\
 &&&1 \\
 &&1&
\end{array}\right).
\end{eqnarray}
With (\ref{boundary-condition-01}), 
the boundary transition matrices are determined as follows:
\begin{eqnarray}
P^L&=\left(\begin{array}{cccc}
 1/N_L&&1/N_L& \\
 &\alpha&&1-\beta \\
 e^{1/T_L}/N_L&&e^{1/T_L}/N_L& \\
 &1-\alpha&&\beta
\end{array}\right),\nonumber\\
P^R&=\left(\begin{array}{cccc}
 1/N_R&1/N_R&& \\
 e^{-1/T_R}/N_R&e^{-1/T_R}/N_R&& \\
 &&\gamma&1-\delta \\
 &&1-\gamma&\delta
\end{array}\right).
\label{heatbath-boundary-matrix}
\end{eqnarray}
Since the bulk transition matrix remains unchanged, 
we only have to replace the parameters in the previous section as follows
\begin{eqnarray*}
q_1^L,\,1-q_2^L\rightarrow 1/N_L ,\qquad 1-q_1^L,\,q_2^L\rightarrow e^{1/T_L}/N_L, \\
q_1^R,\,1-q_2^R\rightarrow 1/N_R, \qquad 1-q_1^R,\,q_2^R\rightarrow e^{-1/T_R}/N_R .
\end{eqnarray*}
The reduced NESS equations are immediately obtained:
\begin{eqnarray}
  1&=( r_1+r_2)/N_R,\nonumber \\
 r_2&=r_1[(1-\gamma)r_7+\delta r_8] ,\nonumber\\
 r_3&=r_1[\gamma r_7+(1-\delta)r_8] ,\nonumber\\
 r_4&=r_1(r_1+r_2)e^{-1/T_R}/N_R ,\nonumber\\
 r_5&=r_1[\gamma r_3+(1-\delta)r_4]/y_3 ,\nonumber\\
 r_6&=r_1[(1-\gamma)r_3+\delta r_4]/y_3 ,\nonumber\\
 r_7&=r_1(r_5+r_6)/(y_1y_3 N_R) ,\nonumber\\
 r_8&=r_1(r_5+r_6)e^{-1/T_R} /(y_1y_3 N_R),
\label{reduced-NESSeq-heatbath-Rsector}
\end{eqnarray}
\begin{eqnarray}
 r_1&=(1+l_7y_1)/N_L ,\nonumber\\
 r_1&=(l_4+l_8)/N_L ,\nonumber\\
 r_1l_3&=y_3[\alpha +(1-\beta)l_6] ,\nonumber\\
 r_1l_4&=y_3[\alpha l_3y_1+(1-\beta)l_5] ,\nonumber\\
 r_1l_5&=(1+l_7y_1)e^{1/T_L}/N_L ,\nonumber\\
 r_1l_6&=(l_4+l_8)e^{1/T_L}/N_L ,\nonumber\\
 r_1l_7&=y_3[1-\alpha+\beta l_6] ,\nonumber\\
 r_1l_8&=y_3[(1-\alpha)l_3y_1+\beta l_5],
\label{reduced-NESSeq-heatbath-Lsector}
\end{eqnarray}
\begin{equation}
 r_7=r_1/y_3.
 \label{reduced-NESSeq-heatbath-Ex}
\end{equation}
These new reduced NESS equations 
(\ref{reduced-NESSeq-heatbath-Rsector})-(\ref{reduced-NESSeq-heatbath-Ex}) 
are solved exactly.  Again, we obtain 
the transfer matrices $X$ and $X'$ of the form (\ref{Transfer-matrices-obtained}) with 
\begin{equation}
 \xi:=\frac1{y_3}= \frac{1-\lambda\mu+(1-\mu) e^{1/T_L}}
{[1-\lambda\mu+(1-\lambda) e^{-1/T_R}]^2}
\Bigl(\lambda e^{-1/T_R}+e^{1/T_L}(1+e^{-1/T_R})\Bigr),
 \label{NESS-heatbath-sol-xi}
\end{equation}
\begin{equation}
 \omega:=y_1y_3
=\frac{1-\lambda\mu+(1-\lambda) e^{-1/T_R}}{[1-\lambda\mu+(1-\mu) e^{1/T_L}]^2}
\Bigl(\mu e^{1/T_L}+e^{-1/T_R}(1+e^{1/T_L})\Bigr),
  \label{NESS-heatbath-sol-omega}
\end{equation}
and the matrices $L$ and $R$ with parameters 
\begin{eqnarray}
 r_1&=\frac{1-\lambda\mu+(1-\lambda) e^{-1/T_R}}
{1-\lambda\mu+(1-\mu) e^{1/T_L}} ,\nonumber\\
 r_2&=(1-\mu)\frac{\lambda e^{-1/T_R}+e^{1/T_L}(1+e^{-1/T_R})}{1-\lambda\mu+(1-\mu)e^{1/T_L}} ,\nonumber\\
 r_3&= r_1^2\xi(e^{-1/T_R}+\mu),\nonumber\\
 r_4&= r_1e^{-1/T_R} ,\nonumber\\
 r_5&= r_1\xi[\gamma r_3+(1-\delta)r_4] ,\nonumber\\
 r_6&=r_1\xi[(1-\gamma)r_3+\delta r_4] ,\nonumber\\
 r_7&= r_1 \xi,\nonumber\\
 r_8&=r_1\xi e^{-1/T_R},
 \label{NESS-heatbath-sol-R}
\end{eqnarray}
\begin{eqnarray}
 l_3&= \frac{e^{1/T_L}+\lambda}{r_1\xi} ,\nonumber\\
 l_4&= \frac{\alpha l_3\xi\omega +(1-\beta)e^{1/T_L}}{r_1\xi} ,\nonumber\\
 l_5&= l_6=e^{1/T_L} ,\nonumber\\
 l_7&= \frac{1-\lambda}{r_1\xi} ,\nonumber\\
 l_8&= \frac{(1-\alpha)l_3\xi\omega+\beta e^{1/T_L}}{r_1\xi},
 \label{NESS-heatbath-sol-L}
\end{eqnarray}
where $\lambda:=\alpha-\beta e^{1/T_L}$ and $\mu:=\gamma-\delta e^{-1/T_R}$.
It is noticeable that, unlike the previous section, the exact solution is 
obtained without any conditions on the six parameters, 
$\alpha,\,\beta,\,\gamma,\,\delta,\,T_L,\,T_R$. 
In the high temperature limit $T_L=T_R=\infty$, it agrees with the result of 
\cite{Prosen16}.

\subsection{Equilibrium state}

When $\alpha=\beta=\gamma=\delta=0$ and $T_L=T_R=T$, 
the system is in contact with heat bath at temperature $T$ only. 
Then,
the steady state of the system is considered to be an equilibrium state at temperature $T$.
We explicitly calculate the probability vector $p_{\bi{s}}$ in this case. 
First, each tensor of PSA is as follows:
\begin{equation}
 X=\left(\begin{array}{cccc}
 1&1&1&1 \\
1 &1&e^{-2/T}&e^{-2/T} \\
1&1&1&1 \\
e^{2/T} &e^{2/T}&1&1
\end{array}\right),
\label{equilibriumX}
\end{equation}
\begin{equation}
 L=\left(\begin{array}{cccc}
 1&1&1&1 \\
e^{1/T}&e^{1/T}&e^{-1/T}&e^{-1/T}
\end{array}\right),\qquad
 R=\left(\begin{array}{cc}
 e^{-1/T} & 1 \\
 e^{-1/T} & e^{-2/T} \\
 e^{-1/T} & 1 \\
 e^{1/T} & 1
\end{array}\right).
\end{equation}
These can be written in the following form
\begin{eqnarray}
 X_{s_2s_3s_4s_5}&=\exp\Bigl[2s_3(s_2-s_4)/T\Bigr] ,\nonumber\\
 L_{s_1s_2s_3}&=\exp\Bigl[s_1(1-2s_2)/T\Bigr] ,\nonumber\\
 R_{s_{n-2}s_{n-1}s_n}&=\exp\Bigl[\Bigl(2s_{n-2}s_{n-1}-2s_{n-1}s_n-1+s_n\Bigr)/T\Bigr].
\end{eqnarray}
Substituting the three expressions to PSA (\ref{PSA01}) and rearranging them, we obtain 
\[
p_{\bi{s}}=\exp\Bigl[\Bigl(s_1+s_n-1+2\sum_{i=1}^{n-1}(-1)^{k}s_is_{i+1}\Bigr)/T\Bigr] .
\]
Because $\sum_{i}E_i=s_1+s_n+\sum_{i=1}^{n-1}2(-1)^is_is_{i+1}$,
the above probability vector represents the equilibrium state at $T$, as expected.

\subsection{Physical quantities in the NESS}
\label{heat-bath-NESS-ev}

\subsubsection*{Partition function: }

The exact solution (\ref{NESS-heatbath-sol-xi})-(\ref{NESS-heatbath-sol-L}) satisfies the eigenvalue equation, 
\begin{eqnarray}
  \bi{l}X=\tau_1 \bi{l}\qquad X\bi{r}=\tau_1\bi{r} ,\nonumber\\
  \bi{l}'X'=\tau_1 \bi{l}'\qquad X'\bi{r}'=\tau_1\bi{r}'  ,
\label{eigenvector-NESS-heatbath}
\end{eqnarray}
whwre
\begin{equation}
 \tau_1=\frac{[\lambda\mu-(1+e^{1/T_L})(1+e^{-1/T_R})]^2}
{[1-\lambda\mu+(1-\mu) e^{1/T_L}][1-\lambda\mu+(1-\lambda) e^{-1/T_R}]}.
\label{tau1}
\end{equation}
Then the partition functions are
\begin{equation}
 Z_n=\tau_1^{n/2-2}\bi{l}\cdot\bi{r}   , \qquad Z'_n=\tau_1^{n/2-2}\bi{l}'\cdot\bi{r}'   .
\end{equation}
From direct calculation, we find
\begin{eqnarray}\fl
  \bi{l}\cdot\bi{r}  =\bi{l}'\cdot\bi{r}' 
 = \frac{(1+e^{1/T_L})(1+e^{-1/T_R})}{1-\lambda\mu+(1-\mu) e^{1/T_L}}
\left[1-\lambda\mu+(\lambda+2) e^{-1/T_R}+(\mu+2) e^{1/T_L}+3e^{\Delta\beta}\right]. 
\label{lr}
\end{eqnarray}
where  $\Delta\beta=1/T_L-1/T_R$\footnote{Do not confuse inverse temperature difference $\Delta\beta$ with left side absorption rate $\beta$.},
and then $Z_n=Z'_n$.

\subsubsection*{Density: }

\begin{figure}[ht]
	\begin{center}
		\includegraphics[scale=0.7]{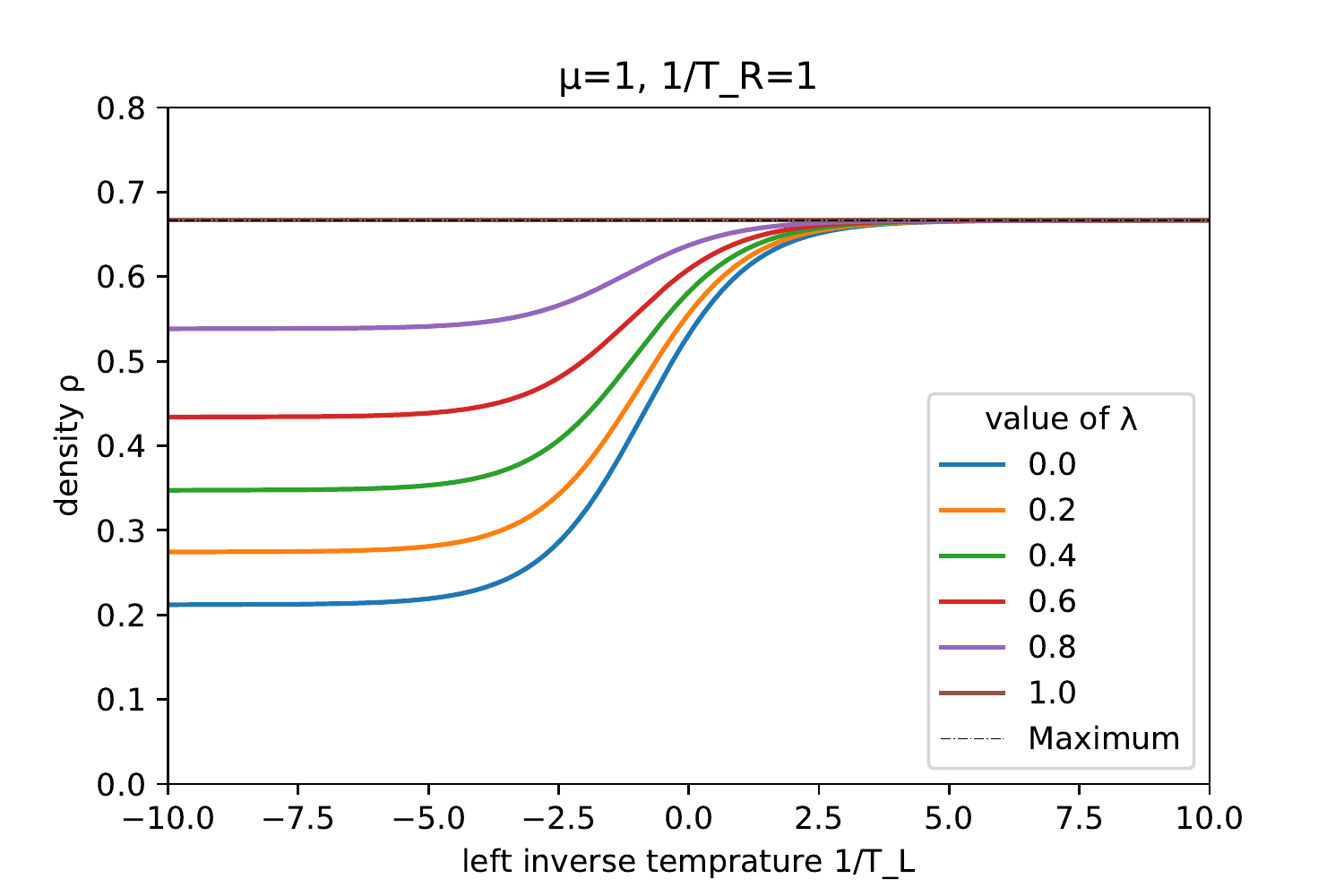}
		\includegraphics[scale=0.7]{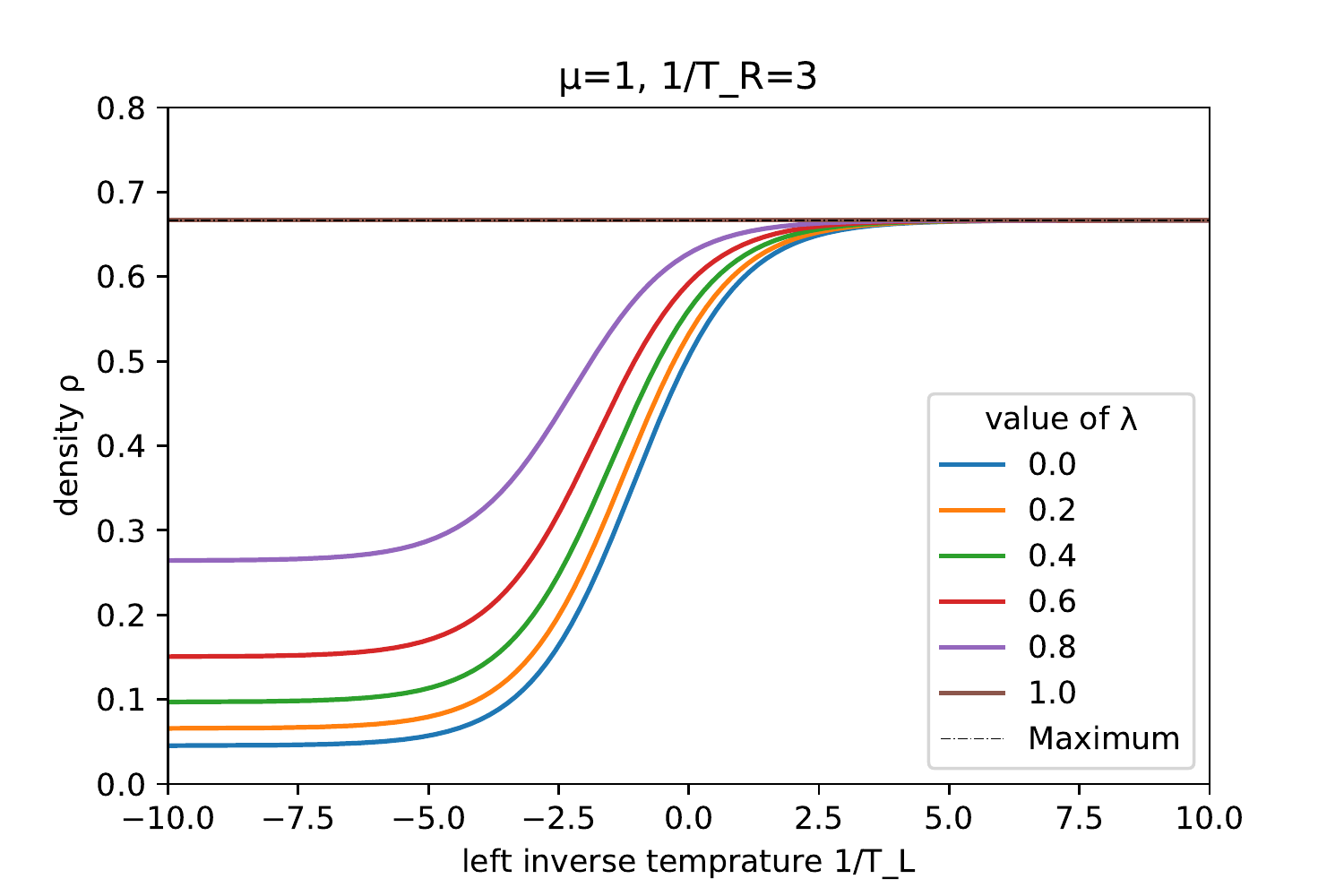}	
	\end{center}
	\caption{In the case of $\mu=1.0$ and (upper)$1/T_R=1.0$ or (lower)$1/T_R=3.0$, and for different values of $\lambda$, we examine the behavior of the bulk density $\rho$ as a function of the inverse temperature $1/T_L$ .
	The maximum density value $\rho=2/3$ is represented by a dotted line, overlapping the $\lambda=1.0$, (brown) line. 
	Indeed, it can be seen that $\rho = 2/3$ regardless of temperature when $\lambda,\,\mu=1$. 
	It can also be seen that as $1/T_L$ decreases and $1/T_R$ increases, $\rho$ gets smaller and asymptotically approaches to $0$. 
	}\label{density}
\end{figure}

Substituting the exact solution into density definition (\ref{RCA54-density}), we obtain
\begin{equation}
\rho_i=\frac{(\lambda+1) e^{-1/T_R}+(\mu+1) e^{1/T_L}+2e^{\Delta\beta}}
{1-\lambda\mu+(\lambda+2) e^{-1/T_R}+(\mu+2) e^{1/T_L}+3e^{\Delta\beta}}.
\label{eq-density}
\end{equation}
for $2\le i\le n-1$. Thus, the density is uniform in this range and the value is denoted 
as $\rho$ in the following. For $i=1$ and $N$, $\rho_i$ can take values different from 
$\rho$, but we do not show them because the equations are cumbersome.
The temperature dependence of the bulk density is illustrated in Figure \ref{density}. 
Let us check some limiting behavior.
For the emission limit $\alpha,\, \gamma \to1,\,\beta,\delta\rightarrow 0$ ($\lambda,\,\mu\to1$), the density takes the maximum value $2/3$ regardless of temperature parameters.  
For the absorption limit $\alpha,\,\gamma \to0,\,\beta,\delta\rightarrow 1$ ($\lambda\to -e^{1/T_L},\,\mu\to -e^{-1/T_R}$), it become
\[
\rho=\frac{x}{2x+1},\qquad x=e^{1/T_L}+e^{-1/T_R},
\]
and is in the range $0<\rho\leq 1/2$. 
$\rho=0$ is established when $1/T_L\to-\infty,\,1/T_R\to\infty$.

In the high temperature limit $T_L,T_R\gg 1$, up to second order of inverse temperatures,
\begin{eqnarray*}
 \rho\sim&\frac{\lambda_0+\mu_0+4}{\lambda_0+\mu_0+8-\lambda_0\mu_0} -\frac{1}{(\lambda_0+\mu_0+8-\lambda_0\mu_0)^2} \\
&\times \Bigl[(\mu_0+2)(\alpha\mu_0+2\alpha-\lambda_0-2)\frac{1}{T_L}-(\lambda_0+2)(\gamma\lambda_0+2\gamma-\mu_0-2)\frac{1}{T_R}\Bigr].
\end{eqnarray*}
Here we have $\lambda_0=\alpha-\beta,\,\mu_0=\gamma-\delta$. 
Furthermore, when the rates of absorption and emission on both boundaries are equal ($\alpha=\gamma,\, \beta=\delta $), it becomes
\begin{equation}
 \rho\sim\frac{2}{4-\lambda_0}+\frac{1-\alpha}{(4-\lambda_0)^2}\Delta\beta,
 \label{hightemp-absemieq-density}
\end{equation}
and depends on the difference of the inverse temperature.

Finally, in the heat conduction limit $\alpha,\beta,\gamma,\delta\to 0$, we have
\begin{equation}
 \rho_i=\frac{e^{1/T_L}+e^{-1/T_R}+2e^{\Delta\beta}}{1+2e^{1/T_L}+2e^{-1/T_R}+3e^{\Delta\beta}}
\end{equation}
for $1\le i\le N$. and in equilibrium $T_L=T_R$, $\rho_i=1/2$.

\subsubsection*{Current: }

\begin{figure}[ht]
	\begin{center}
		\includegraphics[scale=0.7]{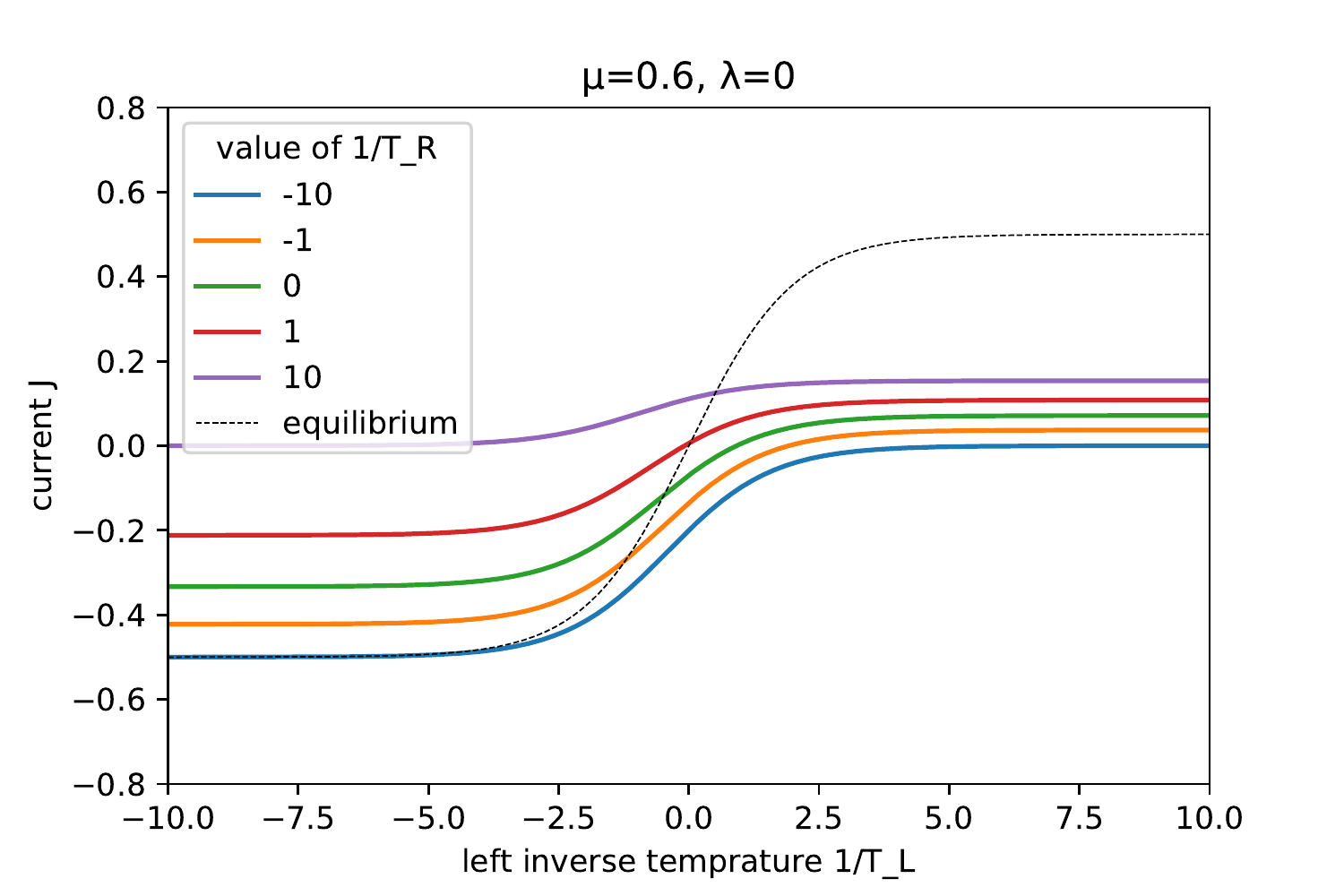}
		\includegraphics[scale=0.7]{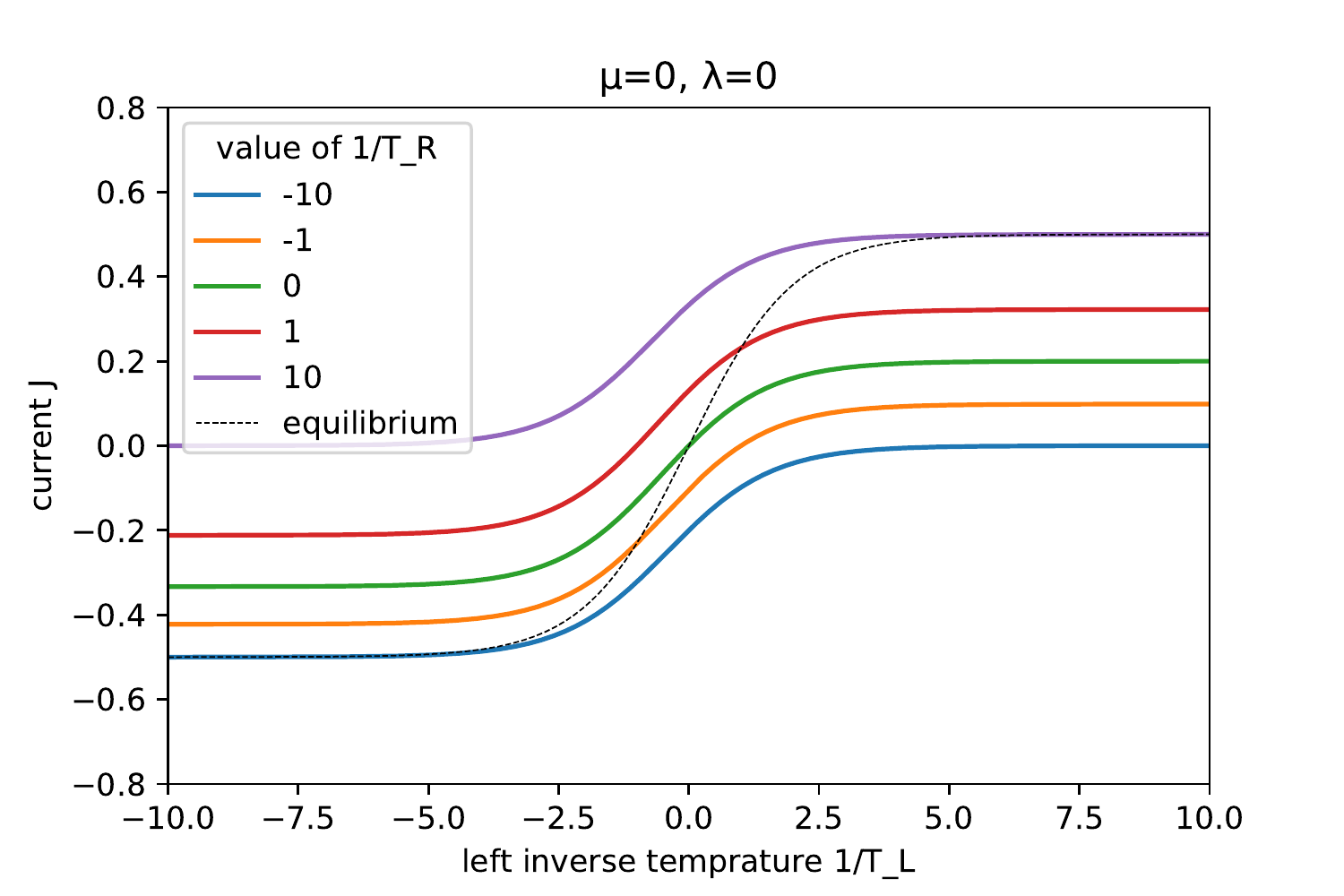}	
	\end{center}
	\caption{In the case of $\lambda=0$, (upper)$\mu=0.6$ or (lower)$\mu=0$, and for different values of $1/T_R$, we examine the behavior of the bulk current $J$ as a function of the inverse temperature $1/T_L$. 
	The equilibrium current with $\lambda,\mu=0$ and $T_L=T_R$ is represented by a dashed line. 
	Even in the equilibrium state, the particle current generally has a non-zero value as a function of temperature.}
	\label{current}
\end{figure}

Substituting the exact solution into the definition of currents (\ref{RCA54-current}), 
we obtain
\begin{eqnarray}
 J_R&= \frac{\lambda e^{-1/T_R}+e^{1/T_L}+e^{\Delta\beta}}
{1-\lambda\mu+(\lambda+2) e^{-1/T_R}+(\mu+2) e^{1/T_L}+3e^{\Delta\beta}}
,\nonumber\\
 J_L&= \frac{\mu e^{1/T_L}+e^{-1/T_R}+e^{\Delta\beta}}
{1-\lambda\mu+(\lambda+2) e^{-1/T_R}+(\mu+2) e^{1/T_L}+3e^{\Delta\beta}}.
 \label{heat54-currentRL}
\end{eqnarray}
The total current is $J=J_R-J_L$. 
The behavior of the current is illustrated in Figure  \ref{current}. 
For the emission limit $\lambda,\,\mu\to1 $, 
the currents become $J_R=1/3,\,J_L=1/3,\,J=0$ regardless of temperature. 
For the absorption limit $\lambda\to -e^{1/T_L},\,\mu\to -e^{-1/T_R}$, it becomes
\begin{equation}
 J=\frac{e^{1/T_L}-e^{-1/T_R}}{1+2(e^{1/T_L}+e^{-1/T_R})}.
\end{equation}
In the heat conduction limit, we have
\begin{equation}
 J=\frac{e^{1/T_L}-e^{-1/T_R}}{1+2(e^{1/T_L}+e^{-1/T_R})+3e^{\Delta\beta}}.
\end{equation}
It is remarkable that the current does not vanish in equilibrium ($T_L=T_R=T$)
but has a finite value $J=\frac{1}{2}\tanh\frac{1}{2T}$. 

\subsubsection*{Energy: }

Let us calculate the expectation values of energy and energy flux using the exact 
NESS solution (\ref{NESS-heatbath-sol-xi})-(\ref{NESS-heatbath-sol-L}). 
Because $|s_i-s_{i+1}|=s_i+s_{i+1}-2s_is_{i+1}$,
the expectation value of energy is given as
\[
 \langle E_{2k}\rangle=2\rho-2J_R ,\qquad 
 \langle E_{2k+1}\rangle=-2\rho+2J_L.
\]
Because the identity $J_L+J_R=\rho$ holds also in this case, we can write 
$\langle E_{2k}\rangle=2J_L$ and $\langle E_{2k+1}\rangle=-2J_R$. 
The total bulk energy is
\begin{equation}
 E=\sum_{k=1}^{n/2-1}(\langle E_{2k}\rangle+\langle E_{2k+1}\rangle)=-(n-2)J.
\end{equation}

\begin{figure}[ht]
 \begin{center}
  \includegraphics[clip,scale=0.8]{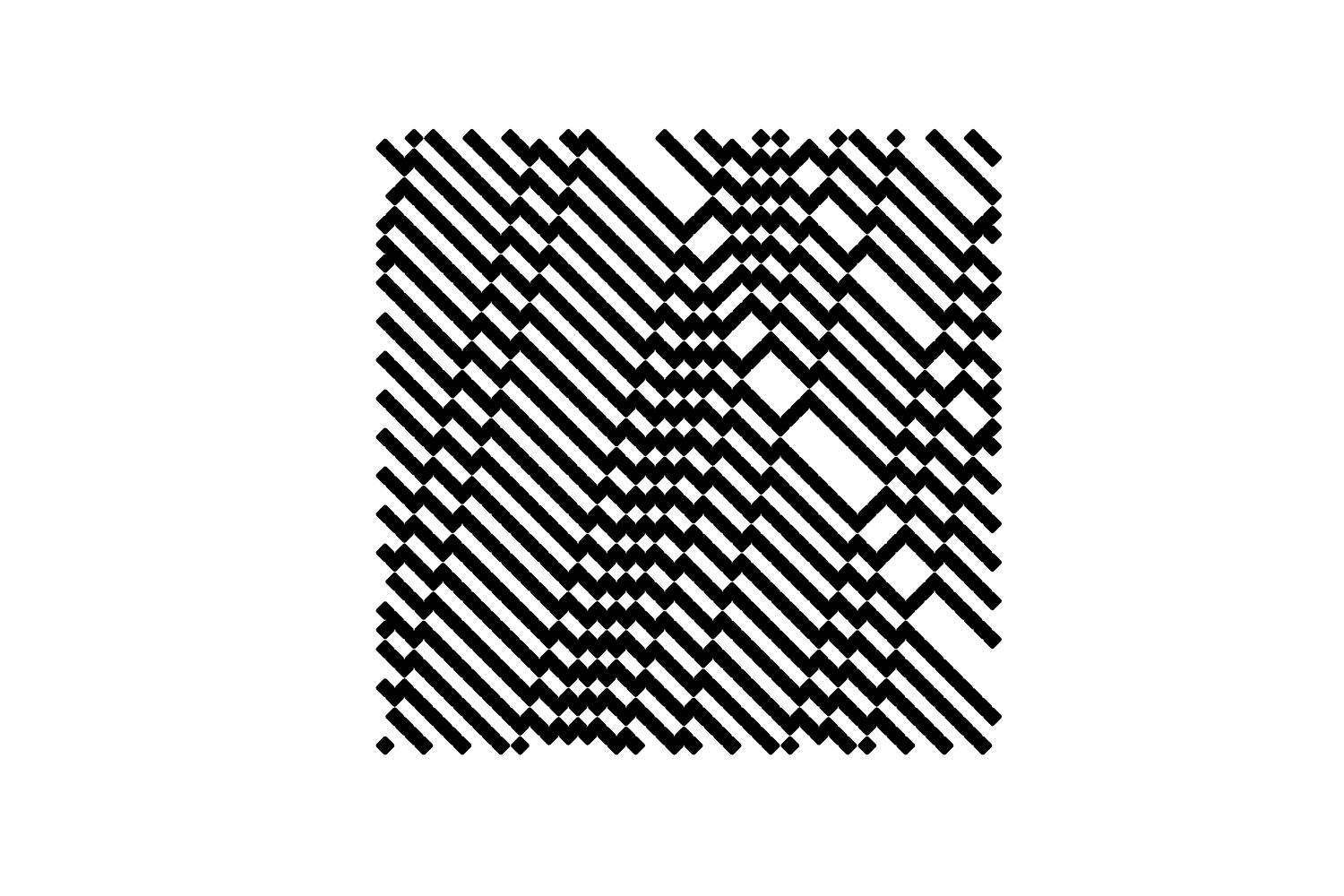}
 \end{center}
\caption{A nonequilibrium steady state without energy flow. Parameter values are
$(\alpha,\beta,\gamma,\delta)=(0.8,0.2,0.1875,0.75)$, $e^{1/T_L}=4.0$, and
$e^{-1/T_R}=0.25$. }
\label{State_without_energyflow}
\end{figure}

Since the NESS vector does not depend on time, we can see immediately that 
the expectation value of energy flux is 
\begin{equation}
  \langle S_i\rangle=1-2\rho=\frac{1-(\lambda+e^{1/T_L})(\mu+e^{-1/T_R})}
{1-\lambda\mu+(\lambda+2) e^{-1/T_R}+(\mu+2) e^{1/T_L}+3e^{\Delta\beta}}.
\end{equation}
Thus, if the condition $(\lambda+e^{1/T_L})(\mu+e^{-1/T_R})=1$ or
\begin{equation}
 [\alpha+(1-\beta)e^{1/T_L}][\gamma+(1-\delta)e^{-1/T_R}]=1
\label{eq-vanishingenergycurrent}
\end{equation}
holds, the energy current vanishes and the bulk density of 1 is 1/2.
Under this condition, the maitrices $X$, $R$, and $L$  are
\begin{equation}
 X=\left(
\begin{array}{cccc}
 1 & 1 & 1 & 1 \\
 1 & 1 & \phi^{-2} & \phi^{-2} \\
 1 & 1 & 1 & 1 \\
 \phi^2 & \phi^2 & 1 & 1
\end{array}
\right)
\end{equation}
\begin{equation}
 L=\left(
\begin{array}{cccc}
 1 & 1 & 1  & 1 \\
\phi-\lambda & \phi-\lambda & 1-\lambda & 1-\lambda
\end{array}
\right),
\end{equation}
\begin{equation}
R=\left(
\begin{array}{cc}
 \phi^{-1} & 1-\mu \\
 \phi^{-1} & \phi^{-2}(1-\mu\phi) \\
 \phi^{-1} & 1-\mu \\
 \phi & 1-\mu\phi
\end{array}
\right)
\end{equation}
where $\phi :=\lambda+e^{1/T_L}$. Thus we can represent the solution with 
only three parameters $\phi$, $\lambda$ and $\mu$. 
Note that $X$, the first row of $L$, and the first column of $R$ are the same as those
of the equilibrium state at $T=(\log\phi)^{-1}$.
Parameters in the partition function,
$\bi{l}\cdot\bi{r}$ and $\tau_1$ are also simplified to
\begin{equation}
 \bi{l}\cdot\bi{r}=\frac{2N_LN_R(1+\phi)}{\phi},\quad
 \tau_1=\frac{(1+\phi)^2}{\phi}
\end{equation}
The densities at the boundary cells are obtained as
\begin{equation}
 \rho_1=\frac{2N_L-1-\phi}{2N_L},\quad \rho_n=\frac{2N_R-1-\phi^{-1}}{2N_R}.
\end{equation}
The currents $J_L$ and $J_R$ are
\begin{equation}
 J_L=\frac{1}{2(1+\phi)}, \quad J_R=\frac{\phi}{2(1+\phi)}, \quad
 J=\frac{\phi-1}{2(1+\phi)}.
\end{equation}
Thus the steady states in this case is the same as the equilibrium one at
temperature $T=(\log \phi)^{-1}$ except the density at the boundary cells.
Such a state is illustrated in Figure \ref{State_without_energyflow}, where we see
more lines downward to the right than those to the left.
In the limit $\lambda,\mu\to 0$, the solutions approach the equilibrium ones.

In the high temperature limit, 
if the absorption and release rates at both boundaries are equal, 
the energy flux become
\begin{equation}
 \langle S_i\rangle\sim\frac{\lambda_0}{4-\lambda_0}-\frac{2(1-\alpha)}{(4-\lambda_0)^2}\Delta\beta,
\end{equation}
and depend only on the temperature difference of the thermal baths at both ends.
In particular, 
if the emission and absorption rates are equal ($\alpha=\beta$),  
we see that $\langle S_i\rangle\sim (1-\alpha)(T_L-T_R)/(8T_LT_R) $.

\section{Discussion}
We have derived two generalizations of Prosen and Mej\'{i}a-Monasterio's result
on nonequilibrium steady states of RCA54.
One is obtained by extending the probabilities for the states of the stochastic cells
at the boundaries. In \cite{Prosen16}, the cells take values 0 or 1 with probability $1/2$.
We have generalized it to $\zeta$ and $1-\zeta$ for the left boundary and 
$\eta$ and $1-\eta$ for the right boundary.
The patch state ansatz has been successfully applied to construct nonequilibrium 
steady states on the 
assumption (\ref{eq-solvability}) for the absorption and emission rates.
The other is obtained by regarding an additive conserved quantity as energy
and employing Boltzmann weights as the probabilities for the stochastic cells.
By doing so, heat-bath temperatures are introduced to the model.
The exact solution in this case has been derived for any set of parameters
without any additional conditions.

Both solutions exhibit uniform density of 1 and ballistic transport.
In the first solution, the current of particles is approximately proportional to
$\gamma-\alpha$ if $\eta=\zeta$ and difference between $\alpha$ and $\gamma$ is 
small.  The second solution has a richer structure, where there exist heat-bath 
temperature and the energy current besides the particle current. 
We have discussed density profile, particle current, energy and energy current
in a variety of limiting cases.   The particle current exists even in equilibrium states,
where energy current vanishes.  It is no wonder because the number of particles
is not a conserved quantity in this system. 
We have explicitly derived the condition for vanishing energy current.
Interestingly, it contains the case with finite emission and absorption rates, where
only the end cells 1 and $n$ show density different from equilibrium state at
effective temperature $(\log\phi)^{-1}$. 

The present method may be further extended to other CA rules.
In ERCA, there are a number of rules that show various types of nonequilibrium 
steady states.  Some rules show flat density profile and ballistic transport like
RCA54, some others show nonflat density profile and still ballistic transport, and 
some others show diffusive behavior with respect to the additive conserved 
quantities. It will be interesting to examine whether the present method can be
generalized to such rules.  It is a future problem.

\end{document}